\documentclass[english,reprint,aps,pre]{revtex4-1}
\usepackage[T1]{fontenc}
\usepackage[latin9]{inputenc}
\usepackage{amsmath}
\usepackage{amssymb}
\usepackage{graphicx}

\makeatletter

\usepackage{babel}

\makeatother

\usepackage{babel}
\begin{document}
\title{Entropy non-conservation and boundary conditions for Hamiltonian dynamical systems}
\date{\today}
\author{Gerard McCaul}
\email{gerard.mccaul@kcl.ac.uk}
\affiliation{Tulane University, New Orleans, LA 70118, USA}
\affiliation{King's College London, London WC2R 2LS, U.K.}
\author{Alexander Pechen}
\email{apechen@gmail.com}

\affiliation{Steklov Mathematical Institute of Russian Academy of Sciences, Gubkina
str. 8, Moscow 119991, Russia}
\affiliation{National University of Science and Technology "MISIS", Leninski
prosp. 4, Moscow 119049, Russia}
\author{Denys I. Bondar}
\email{dbondar@tulane.edu}

\affiliation{Tulane University, New Orleans, LA 70118, USA}

\begin{abstract}
Applying the theory of self-adjoint extensions of Hermitian operators
to Koopman von Neumann classical mechanics, the most general set
of probability distributions is found for which entropy is conserved
by Hamiltonian evolution. A new dynamical phase associated
with such a construction is identified. By choosing distributions
not belonging to this class, we produce explicit examples of both free
particles and harmonic systems evolving in a bounded phase-space in such a way that entropy is nonconserved. While these nonconserving states are classically forbidden, they may be interpreted as states of a quantum system tunneling through a potential barrier boundary. In this case, the allowed boundary conditions are the only distinction between classical and quantum systems. We show that the boundary conditions for a tunneling quantum system become the criteria for entropy preservation in the classical limit. These findings highlight how boundary effects drastically change the nature of a system.
\end{abstract}
\maketitle

\section{Introduction}

Entropy is a critical element of physical theories, defining the notion of equilibrium \citep{Lieb2000},
as well as serving as a measure of irreversibility \citep{Ford2015} and the arrow of time
\citep{1367-2630-11-7-073008,Timearrow,Batalhao2015}. In thermodynamics, the physical content of entropy is contained in its time dynamics, rather
than its absolute value. Given the vital role which entropy plays
in characterizing physical systems, the conditions under which entropy
is time dependent are of great interest.

While entropy is an integral component of our understanding of thermodynamics and information, it is not a microscopic quantity and its definition therefore contains a degree of flexibility \citep{Hanggi}. The entropy of a system depends on the level of course graining in the phase
space, and on the variables considered \citep{Jaynes1992}.
In this paper we shall adopt the Gibbs measure of entropy for a probability
density $\rho$ of a  one-dimensional system with coordinate $q$ and momentum $p$,
\begin{equation}
S=-\int{\rm d}q{\rm d}p\rho\ln\rho\label{eq:entropy}
\end{equation}
which at equilibrium is equivalent to the thermodynamic entropy \citep{RevModPhys.61.981}.

It is often stated that, for a reversible process, the entropy change
is zero. Another common assumption is that Hamiltonian
systems are time reversible.
The two ideas are often combined
into the claim that systems with a Hamiltonian evolution conserve
entropy. This is not in fact the case.  The probability density $\rho$ of a Hamiltonian system evolves according to the Liouville equation
\begin{equation}
\dot{\rho}=\left\{ H,\rho\right\}. \label{eq:Liouvilleevolution}
\end{equation}
Taking the time derivative of Eq. (\ref{eq:entropy}), substituting the Poisson bracket $\{\cdot,\cdot\}$,  and integrating this by parts, one finds
that the entropy production arises solely from the boundary terms. For a system with box boundaries at $q_{\pm}$, $p_{\pm}$,
\begin{eqnarray}
\dot{S} & = & \left[\int_{q_{-}}^{q_{+}}{\rm d}q\rho\frac{\partial^{2}H}{\partial q\partial p}\right]_{p_{-}}^{p_{+}}-\left[\int_{p_{-}}^{p_{+}}{\rm d}p\rho\frac{\partial^{2}H}{\partial q\partial p}\right]_{q_{-}}^{q_{+}}\nonumber \\
 &  & +\left[\int_{p_{-}}^{p_{+}}{\rm d}p\rho\left(\ln\rho+1\right)\frac{\partial H}{\partial p}\right]_{q_{-}}^{q_{+}}\nonumber \\
 &  & -\left[\int_{q_{-}}^{q_{+}}{\rm d}q\rho\left(\ln\rho+1\right)\frac{\partial H}{\partial q}\right]_{p_{-}}^{p+} \label{entropyrate}
\end{eqnarray}
It is at this point that one generally assumes boundary conditions
to kill these terms \citep{LandauLifshitz}, but this assumption is overly restrictive,
and peculiar to both the dynamics and distribution which is being evolved.
This assumption does not give a \emph{ general} answer to the question of entropy conservation.
At the same time, it is difficult to assess how modifications, such
as a restriction in phase space, would affect the entropy production.

To answer the question of entropy conservation with the greatest
possible generality, we shall employ the Koopman von-Neumann (KvN)
formalism. In particular, we shall use this formalism to reduce the problem of entropy
conservation to the analysis of a classical self-adjoint evolution operator.

The theory of self-adjoint extensions of Hermitian operators is closely related to boundary conditions, and is vital in
the resolution of apparent paradoxes in quantum mechanics \citep{Bonneau2001}.
The theory also makes explicit connection to topological phases of matter \citep{Chen2014,Chen2012,Ahari2016}
and spontaneously broken symmetries \citep{Araujo2004,Berman1991,Capri1977}. 
Moreover, self-adjointness assumes a vital role in novel approaches
to prove the Riemann zeta hypothesis \citep{Bender2017}. It is, in
short, a useful formalism that may be applied to the problem
of entropy conservation when alloyed to the KvN description of classical dynamics.

We proceed as follows: A brief overview of KvN mechanics is provided
in Section II. Section III outlines the theory of self-adjoint operators necessary for our analysis, and the conditions under which a Hermitian operator is self adjoint. The equivalence of the criteria for a generator of time translations to be self-adjoint and for that evolution to conserve entropy is shown. Section IV applies the developed technique to a classical oscillator,
and establishes the conditions for entropy conservation when its coordinate is restricted to the half-line (i.e., a phase space $\mathcal{P} =\mathbb{R}_+ \times \mathbb{R})$. Section V provides specific counterexamples of distributions evolved by Hamiltonian dynamics which violate entropy conservation. Section VI discusses in detail these entropy violating distributions, providing an interpretation of their physical significance and relation to quantum dynamics.
Finally, Sec. VII concludes with a discussion of the results.

\section{Koopman-von Neumann Dynamics \label{sec:KvN dynamics}}

KvN mechanics is the reformulation of classical mechanics in a Hilbert-space formalism \citep{Koopman315}. This operational formalism underlies ergodic theory \citep{ReedSimon2}, and is a natural method for modeling quantum-classical hybrid systems  \citep{Sudarshan1976,Viennot_2018,Bondarquantumclassical}. KvN dynamics has also been applied to establish a classical speed limit for dynamics\citep{PhysRevLett.120.070402} and an alternate formulation of classical electrodynamics \citep{Hilbertelectrodynamics}.  By using the KvN formalism it is even possible to combine classical and quantum dynamics in a unified framework known as operational dynamical modeling \citep{Bondar2012a,Bondar2013}. 

The KvN formalism has been successfully applied to construct both deterministic and stochastic classical path integrals \citep{Gozzi2014, 1505.06391}, including generalizations with geometric forms \citep{Gozzi1994}. These path integral formulations may be usefully applied with classical many-body diagrammatic methods {\citep{Liboff}} and for the derivation and analysis of generalized Langevin equations \citep{Mythesis}. KvN has been used to study specific phenomena, including examinations of dissipative behavior
\citep{Chruscinski2006}, linear representations of nonlinear
dynamics \citep{Koopman-nonlinear}, analysis of the time-dependent harmonic oscillator \citep{RamosPrieto2018}, and other industrial applications \citep{mezic2005spectral,budivsic2012applied}.  

Both KvN and quantum dynamics begin by defining the Hilbert space ${\cal H}=L^{2}\left(\mathcal{P},{\rm d}\mu\right)$:
\begin{equation}
{\cal H}=L^{2}\left(\mathcal{P},{\rm d}\mu\right)=\left\{ \phi\,:\,\mathcal{P}\to\mathbb{C}\,\left|\,\int{\rm d}\mu\left|\phi\right|^{2}<\infty\right.\right\}, 
\end{equation} 
which is the set of all functions on the space $\mathcal{P}$ that
are square integrable with measure ${\rm d}\mu$. In classical KvN dynamics $\mathcal{P}$ is a phase space, while in quantum mechanics it is a configuration space. The inner product of the elements of this Hilbert space is
\begin{equation}
\left\langle \phi\left|\psi\right.\right\rangle =\int{\rm d}\mu\phi^{*}\psi,
\end{equation}
for $\phi,\psi\in \mathcal{H}$. The
elements of ${\cal H}$ are interpreted as (unnormalized) probability
density amplitudes which obey the Born rule in both quantum and KvN
mechanics \citep{BrumerBornRule},
\begin{equation}
\left|\psi \right|^2=\rho.
\end{equation}

In both cases, the dynamics of elements of ${\cal H}$ are given by a one-parameter, strongly continuous group of unitary transformations on $\mathcal{H}$,  
\begin{equation}
    \psi(t)=U(t)\psi(0).
\end{equation}
The form of this family of unitary transformations is, by Stones'
theorem  \citep{Stonetheorem}:
\begin{equation}
U\left(t\right)={\rm e}^{i\hat{A}t},
\end{equation}
where $\hat{A}$ is a unique self-adjoint operator. Self-adjointness is a stronger condition than Hermiticity, and its necessity shall be demonstrated in Sec. \ref{sec:selfadjointoperators}.  The family of unitary
transformations may be interpreted as time evolution, leading to the differential equation: 
\begin{equation}
\dot{\psi}=i\hat{A}\psi.
\end{equation}
Specifying the form of this operator adds physics to the formalism. For example, one may derive both quantum and classical dynamics from this model by defining a fundamental commutation relation and assuming that observable expectations obey the Ehrenfest theorems. 

The sole distinction between Koopman dynamics
and quantum mechanics is in the choice
of commutation relations \citep{Bondar2012a}. In the quantum case $\left[\hat{x},\hat{p}\right]=i\hbar$, and the self-adjoint operator is then $\hat{A}=-\frac{1}{\hbar}\hat{H}$, 
yielding: 
\begin{equation}
i\hbar\dot{\psi}_{{\rm qm}}=\hat{H}\psi_{{\rm qm}},
\end{equation}
 which is the familiar Schr{\"o}dinger equation. In KvN dynamics, $\left[\hat{x},\hat{p}\right]=0$,
and $\hat{A}$ is the Koopman operator $\hat{K}$ defined as
\begin{equation}
\hat{K}\psi_{\rm cl}=-i\left\{ H,\psi_{\rm cl}\right\} .
\end{equation}
This leads to the remarkable conclusion that probability density amplitudes
are governed by the Liouville equation, in the same way as probability
densities themselves: 
\begin{equation}
\dot{\psi}_{{\rm cl}}= i\hat{K}\psi_{\rm cl}=\left\{ H,\psi_{{\rm cl}}\right\}. 
\end{equation}

The choice of commutation relation also
defines the measure on the Hilbert space. While ${\rm d}\mu={\rm d}q{\rm d}p$
is a valid measure in the classical case, only ${\rm d\mu={\rm d}q}$
(in the coordinate representation) preserves non-negativity for the quantum commutation
relation \citep{measure}. One may still use the full phase
space quantum mechanically, but the object under consideration in
this case is the Wigner quasiprobability distribution \citep{Baker1958,Curtright2014,Groenewold1946}.

It may appear at first glance that the KvN framework is a formal hammer
cracking a classical nut, but this formulation provides a key advantage;
one may exploit the deep theory of operators
on Hilbert space. In particular, the theory of self-adjoint operators
allows the same general statements to be made about entropy conservation
as in the quantum case. Explicitly, if $\hat{K}$ is self-adjoint,
then the Gibbs entropy will be conserved. To prove this however, 
we must first review the definition of a self-adjoint operator.

\section{Criteria for Conservation of Entropy}

\subsection{Self-Adjoint Operators \label{sec:selfadjointoperators}}

An operator in the Hilbert space $\mathcal H$ is a pair $(\mathcal{D}(\hat{A}), \hat{A})$ where $\hat{A}$ is a linear mappings of elements in a subset $\mathcal{D}(\hat{A})$ of the Hilbert space onto $\mathcal{H}$, 
\begin{equation}
    \hat{A}: \mathcal{D}(\hat{A}) \to \mathcal{H}.
\end{equation} 
The set $\mathcal{D}(\hat{A})$ is the domain of the operator, i.e., the set of vectors in the Hilbert space for which the operator mapping is defined. If the Hilbert space is finite, operators may be represented as finite matrices, and multiplication of a vector by a matrix is well defined for all vectors in a finite-dimensional Hilbert space, in this case $\mathcal{D}(\hat{A})\equiv \mathcal{H}$~\citep{LinearOperatorsHilbertSpace}. For infinite-dimensional Hilbert spaces however, this is not the case and the operator domain does not necessarily coincide with the full Hilbert space.

Assuming an operator $\hat{A}$  is densely defined [i.e., $\mathcal{D}(\hat{A})$ is a dense subspace of $\mathcal{H}$] \citep{ReedSimon2}, one may define its adjoint operator $\hat{A}^\dagger$ as follows. A vector  $\phi\in\mathcal H$ belongs to the domain $\mathcal{D}(\hat{A^{\dagger}})$ if and only if the linear functional $f_\phi(\psi):=\langle \phi|\hat A\psi\rangle$ is continuous. Then, by the Riesz representation theorem, there exists a unique $z\in\mathcal H$ such that $\langle \phi|\hat A\psi\rangle=\langle z|\psi\rangle$ for any $\psi\in{\cal D}(\hat A)$. Action of the adjoint operator on $\phi$ is defined as $\hat A^\dagger\phi=z$. One has
\begin{equation}
\langle\phi|\hat{A}\psi\rangle=\langle\hat{A}^\dagger\phi|\psi\rangle \quad \forall \phi \in \mathcal{D}(\hat{A^\dagger}),\; \psi \in \mathcal{D}(\hat{A}). 
\end{equation} 
It may happen that the adjoint operator does not exist (its domain may not be dense or even empty; see, e.g., Example 4 in Sec. VIII.1, Vol. 1 of Ref. \citep{ReedSimon2}). This occurs if the graph of the operator; that is, the set of pairs $(\psi, \hat A\psi)$, $\psi\in D(\hat A)$, is not closable in $\mathcal H \times \mathcal H$. Otherwise the adjoint operator is correctly defined.

A \emph{Hermitian} (or \textit{symmetric}) \citep{ReedSimon2} operator is such that $D(\hat A)\subseteq D(\hat A^\dagger)$ and the action of the operator and its adjoint is identical on $\mathcal{D}(A)$. Then
\begin{equation}
\langle\phi|\hat{A}\psi\rangle=\langle\hat{A}\phi|\psi\rangle, \quad \forall \phi, \psi \in \mathcal{D}(\hat{A}). \label{eq:Hermitianconditionoperator}
\end{equation}
In the special case $\mathcal{D}(A)=\mathcal{D}(A^\dagger)$ the operator is called \emph{self-adjoint}. The condition $ \mathcal{D}(\hat{A})\subseteq \mathcal{D}(\hat{A^{\dagger}})$ requires that the adjoint of a symmetric operator is defined for any element in $\mathcal{D}(\hat{A})$, but does not otherwise restrict its domain. Figure 1 illustrates the fact that the domain of a symmetric operator on a Hilbert space is a subset of the domain of its adjoint operator.

To illustrate the difference between a Hermitian and self-adjoint operator, we evaluate a Koopman operator with domain 
\begin{align}
    \mathcal{D}(\hat{K})= & \left\{\psi \left| \psi \in \mathcal{H} \textrm{ such that }  \frac{\partial\psi}{\partial q},\frac{\partial\psi}{\partial p} \in \mathcal{H} \right. \right. \nonumber \\ & \ \  \bigg. \textrm{and } \psi\left(q_{\pm},p\right)=\psi\left(q,p_{\pm}\right)=0 \bigg\}. \label{exampledomain}
\end{align} 
The Hermitian property of this operator may be investigated by using integration by parts:
\begin{eqnarray}
\langle\phi|\hat{K}\psi\rangle & = & i\int{\rm d}q{\rm d}p\ \phi^{*}\left(\frac{\partial H}{\partial p}\frac{\partial\psi}{\partial q}-\frac{\partial H}{\partial q}\frac{\partial\psi}{\partial p}\right)\nonumber \\
 & = & i\left[\int{\rm d}p\ \phi^{*}\psi\frac{\partial H}{\partial p}\right]_{q_{-}}^{q_{+}}-i\left[\int{\rm d}q\ \phi^{*}\psi\frac{\partial H}{\partial q}\right]_{p_{-}}^{p_{+}}\nonumber \\
 &  &+ \langle\hat{K}\phi|\psi\rangle.
\end{eqnarray}
If $\hat{K}$ is Hermitian, the following boundary condition holds: 
\begin{equation}
\left[\int{\rm d}p\ \phi^{*}\psi\frac{\partial H}{\partial p}\right]_{q_{-}}^{q_{+}}-\left[\int{\rm d}q\ \phi^{*}\psi\frac{\partial H}{\partial q}\right]_{p_{-}}^{p_{+}}=0, \label{eq:boundaries self-adjoint}
\end{equation}
which is automatically satisfied by using the example domain of Eq. (\ref{exampledomain}). As a result, these boundary conditions for the wavefunction specify the operator's domain.

For the example domain, the boundary condition is satisfied regardless of the domain of the adjoint operator, $\mathcal{D}(\hat{K^\dagger}) \supseteq \mathcal{D}(\hat{K})$.  This potential mismatch in the domains of an operator and its adjoint is deeply problematic, as it implies a violation of time-reversal invariance \citep{Bonneau2001}. For this reason, the generator of time-translations (and in fact \emph{all} observable operators) must be self-adjoint, as defined above. Given an operator with a defined action, it is a nontrivial exercise to find which domains, if any, will make the operator self-adjoint. Fortunately, there is a powerful theorem of functional analysis that may be applied to this problem.

\subsection{The von Neumann deficiency index theorem}

The \emph{von Neumann deficiency index
theorem} \citep{neumannbook} can be used to answer the question of whether an operator $\hat{A}$ may be made self adjoint. Details of this important theorem may be found in Refs. \citep{Bonneau2001, LinearOperatorsHilbertSpace, naimark2009linear}, but we shall briefly outline its use here. 

The deficiency index theorem determines all possibilities for a Hermitian operator $\hat{A}$ with domain $\mathcal{D}(\hat{A})$ by considering the eigenstates in $\mathcal{D}(\hat{A}^{\dagger})$
with imaginary eigenvalues, i.e., those $\psi_{\pm}\in\mathcal{D}(\hat{A}^{\dagger})$
satisfying the equation: 
\begin{equation}
\hat{A}^{\dagger}\psi_{\pm}=\pm i\psi_{\pm}. \label{deficiencyeigenstates}
\end{equation}
The number of linearly independent solutions
for $\psi_{\pm}$ are the \emph{deficiency indices} $n_{\pm}$. They
determine the following three possibilities \citep{ReedSimon2}:
\begin{eqnarray*}
 & n_{+}=n_{-}=0 & \quad\hat{A}\ \textrm{is essentially self-adjoint.}\\
 & n_{+}=n_{-}=n\geq1 & \quad\hat{A}\ \textrm{has infinite self-adjoint extensions.}\\
 & n_{+}\neq n_{-} & \quad\hat{A}\ \textrm{has no self-adjoint extension.}
\end{eqnarray*}

\begin{figure}
\includegraphics[width=0.5\textwidth]{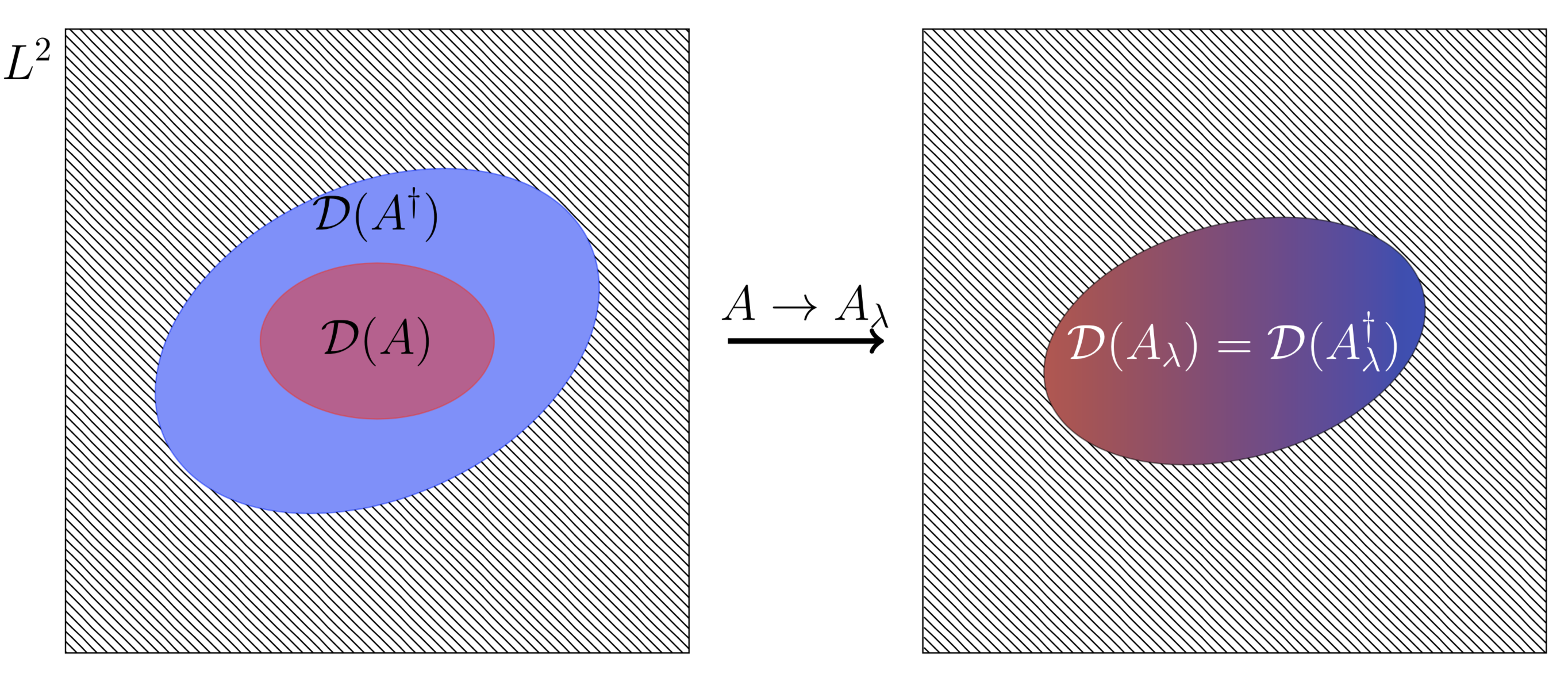}\caption{The domain of a symmetric operator $\hat{A}$ is a subset of the domain of its adjoint operator. A self-adjoint extension modifies the domains of both operators such that they become identical. \label{fig:An-operator-}}
\end{figure}

A self-adjoint extension is an operator $\hat{A}_{\lambda}$
with the same action as $\hat{A}$ on $D(\hat A)$, but whose domain has been modified, $\mathcal{D}(\hat{A})\to\mathcal{D}(\hat{A}_{\lambda})\supseteq D(\hat A)$, to enforce the self-adjoint condition. Figure \ref{fig:An-operator-} shows an example
schematic, demonstrating the domain modification made by a self-adjoint
extension.

If $n_{+}=n_{-}=n$, each self adjoint extension is characterized by $n^2$ parameters, in the form of an $n\times n$ unitary matrix $U$.  For the Koopman operator, the boundary condition corresponding to a self-adjoint extension with domain $\mathcal{D}(\hat{A}_{\lambda})$ is simply Eq. (\ref{eq:boundaries self-adjoint}), replacing $\phi$ with $\left(\psi^n_+ + \sum_m U_{n,m}\psi^m_{-}\right)$ \citep{Araujo2004}:

\begin{align}
\forall n: &\left[\int{\rm d}p\ \left(\psi^n_+ + \sum_m U_{n,m}\psi^m_{-}\right)^{*}\psi\frac{\partial H}{\partial p}\right]_{q_{-}}^{q_{+}}& \nonumber \\ =&\left[\int{\rm d}q\ \left(\psi^n_+ + \sum_m U_{n,m}\psi^m_{-}\right)^{*}\psi\frac{\partial H}{\partial q}\right]_{p_{-}}^{p_{+}}. \label{deficiencyboundarycondition}
\end{align}
Here $\psi_{\pm}^n$ is one of the $n$ solutions to Eq. (\ref{deficiencyeigenstates}) and $U_{n,m}$ is an element of the unitary matrix $U$. Consequently, one may use the deficiency index theorem to characterize the most general boundary condition on $\psi$ for which $\hat{K}$ is self-adjoint in terms of $U$.

\subsection{Self-Adjoint Operators and Entropy Conservation}
While the von Neumann deficiency index theorem establishes the conditions under which $\hat{K}$ is self-adjoint, this does not explicitly address the question of entropy conservation. We now demonstrate that a self-adjoint Koopman operator guarantees 
entropy conservation. First, define the operator \footnote{In the quantum case, the trace of this operator is the von Neumann
entropy. This similarly obeys an $H$ theorem when $\hat{\rho}$ is
evolved by a Lindblad equation \citep{Alicki}.}: 
\begin{eqnarray}
\hat{S}=-\hat{\rho}_{{\rm cl}}\ln\hat{\rho}_{{\rm cl}},
\end{eqnarray}
where 
\begin{equation}
\hat{\rho}_{{\rm cl}}=\left|\psi_{{\rm cl}}\left\rangle \right\langle \psi_{{\rm cl}}\right|.
\end{equation}
Note that the $\hat{S}$ operator is explicitly nonlinear; however, the proof of its time independence does not rely on linearity, as if $\hat{K}$ is self-adjoint, then 
\begin{equation}
{\rm Tr}[\hat{S}\left(t\right)]={\rm Tr}\left[{\rm e}^{-i\hat{K}t}\hat{S}\left(0\right){\rm e}^{i\hat{K}^{\dagger}t}\right]={\rm Tr}[\hat{S}\left(0\right)],
\end{equation}
where in the final equality, we have exploited that $\hat{K}$ is
self-adjoint and all the traces are finite. Having demonstrated the
time independence of the trace of $\hat{S}$, we now express it in
the $q,p$ basis: 
\begin{equation}
\hat{S}=-\int{\rm d}q{\rm d}p\,\rho\left|q,p\left\rangle \right\langle q,p\right|{\rm ln}\left(\rho\int{\rm d}q'{\rm d}p'\left|q',p'\left\rangle \right\langle q',p'\right|\right).
\end{equation}
Expanding this yields: 
\begin{equation}
\hat{S}=-\int{\rm d}q{\rm d}p\,\rho\ln\left(\rho\right)\left|q,p\left\rangle \right\langle q,p\right|,
\end{equation}
and therefore 
\begin{equation}
{\rm Tr}[\hat{S}]=-\int{\rm d}q{\rm d}p\,\rho\ln\rho=S.
\end{equation}
Thus, we establish that ${\rm Tr}[\hat{S}]$ is the entropy, and is
conserved for self-adjoint $\hat{K}$. Note, however, that entropy conservation is not in itself a guarantee of unitary evolutions.

To summarize, entropy conservation is guaranteed when the Koopman operator $\hat{K}$ is self-adjoint. The domain of the self-adjoint extensions may be determined by  using the deficiency index theorem. This domain corresponds to the most general boundary
conditions which conserve entropy for a given Hamiltonian.

Establishing the self-adjoint extensions of an operator is a nontrivial exercise, but the analysis is most straightforward in
a system whose Koopman operator is functionally dependent on only
one phase-space coordinate. Therefore, in the following we consider the specific examples of free and periodic systems in
action-angle coordinates, but emphasize that this is a choice of computational
convenience in applying generic arguments.

\section{Entropy Conservation For Cyclic Systems}
We now apply the results of the previous section to derive the most general entropy-preserving boundary conditions for both the harmonic oscillator and free particle with a bounded phase space. The first section will consider general box boundaries in action-angle coordinates. 

\subsection{Self-adjoint domain In action-angle coordinates}
\begin{figure}
\includegraphics[height=4cm]{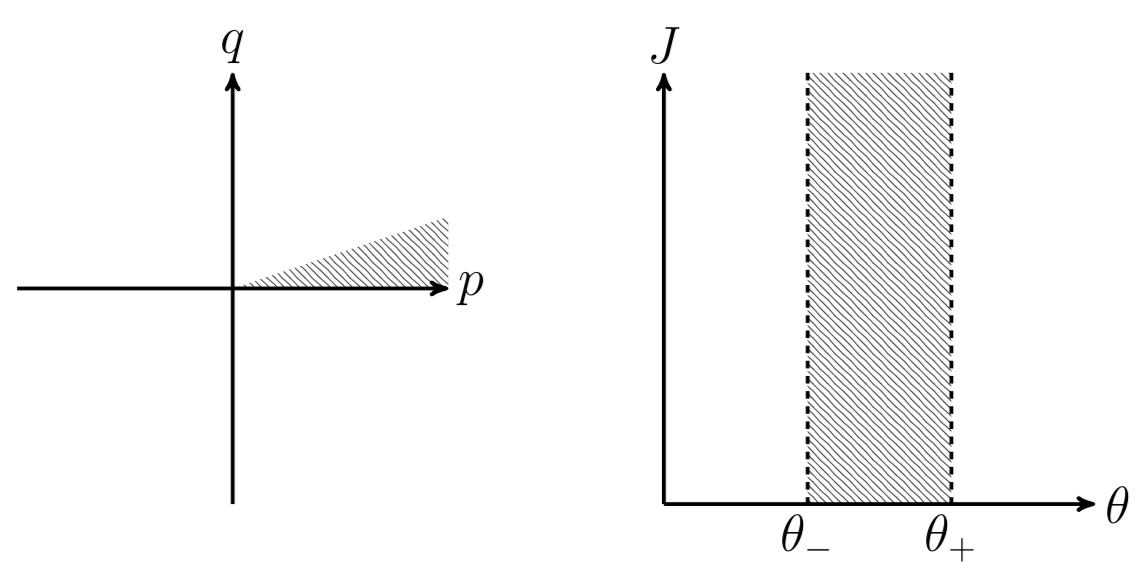} \caption{Example phase-space boundaries in (left) original coordinate system
and (right) action-angle coordinates. \label{fig:Example-phase-space-boundaries}}
\end{figure}

 Consider a system whose coordinates can be canonically transformed
into a space where one coordinate is cyclic, i.e., a Hamiltonian that functionally
depends on only one phase-space coordinate. Any periodic system may be described by these canonical action-angle coordinates \citep{Goldstein}, but
here we take the simplest example of a harmonic oscillator: 
\begin{equation}
H=\frac{\omega^2 q^{2}+p^{2}}{2}.
\end{equation}
Using action-angle coordinates $(\theta, J)$ greatly simplifies the
analysis, and can be achieved with the substitution: 
\begin{equation}
q=\sqrt{\frac{2J}{\omega}}\sin\theta,\quad p=\sqrt{2\omega J}\cos\theta.
\end{equation}
In the action-angle representation, we choose box phase-space boundaries, with $\theta\in\left[\theta_{-},\theta_{+}\right]$
and $J\in\left[0,J_{b}\right]$ (see Fig. \ref{fig:Example-phase-space-boundaries}),
which contains a particle restricted to the half-line $q>0$ as a special case. In the new coordinates,
$H=\omega J$, leading to the following form of the Koopman operator for
the harmonic oscillator: 
\begin{equation}
\hat{K}\psi=-i\left\{ H,\psi\right\} =-i\omega\frac{\partial\psi}{\partial\theta}.
\end{equation}

We now apply the deficiency index theorem in this new basis by finding the wavefunctions satisfying
\begin{equation}
\hat{K}\psi_{\pm}=\pm i\psi_{\pm},
\end{equation}
meaning that we look for solutions to 
\begin{equation}
\frac{\partial\psi_{\pm}}{\partial \theta}=\mp\frac{1}{\omega}\psi_{\pm}.
\end{equation}
Solutions to this equation have the form 
\begin{equation}
\psi_{\pm}(\theta,J)=f_{\pm}(J)e^{\mp \frac{\theta}{\omega}} \label{eq:deficiencysolution}.
\end{equation} 
The solutions which belong to  $L^{2}$ are all those for which:
\begin{align}
\int_{\theta_{-}}^{\theta_{+}}{\rm d}\theta\int_{0}^{J_{b}}{\rm d}J\ \left|\psi_{\pm}\right|^{2}=\int_{0}^{J_{b}}{\rm d}J\ \nonumber \left|f_{\pm}\left(J\right)\right|^{2}\int_{\theta_{-}}^{\theta_{+}}{\rm d}\theta\ {\rm e}^{\mp2\frac{\theta}{\omega}}
\\ =\mp\frac{\omega}{2}\left(\int_{0}^{J_{b}}{\rm d}J\ \left|f_{\pm}\left(J\right)\right|^{2}\right)\left({\rm e}^{\mp2\frac{\theta_+}{\omega}}-{\rm e}^{\mp2\frac{\theta_-}{\omega}}\right) <\infty.
\end{align}
Any function $f_{\pm}$ satisfying
$\int_{0}^{J_{b}}{\rm d}J\ \left|f_{\pm}\left(J\right) \right|^2<\infty$ determines a particular solution to $\psi_{\pm}$. From this we conclude that $n_{+}=n_{-}=\infty$.

Rather than apply Eq. (\ref{deficiencyboundarycondition}) directly, in this case the Koopman operator has the same form as the quantum-mechanical momentum operator on a fixed interval. Following the examples of Refs. \citep{Araujo2004,Bonneau2001} we conclude that for each value of $J$, the Koopman operator has a one parameter self-adjoint extension. Labeling each of these arbitrary parameters by $\beta(J)$,  we find the boundary condition for the self-adjoint domain of $\hat{K}$:
\begin{equation}
\psi\left(\theta_{+},J\right)={\rm e}^{i\beta\left(J\right)}\psi\left(\theta_{-},J\right).\label{eq:selfadjointcondition}
\end{equation} 
While this phase is not directly observable
in expectations, it \emph{is} gauge-invariant (in the sense of locally
rotating a complete set of states). 

Potential physical consequences of the choice of self-adjoint extension can be examined by observing that time evolution corresponds to a translation in $\theta$. Including the time argument explicitly in the wavefunction, harmonic dynamics guarantee
\begin{equation}
\psi\left(\theta,J,t\right)=\psi\left(\theta-\omega t,J,0\right).
\end{equation}
In particular, the wavefunction at the $\theta$ boundaries may be expressed in terms of a time translation of the wavefunction at an arbitrary point in $\theta$
\begin{align}
\psi\left(\theta_{\pm},J,t\right)&=\psi\left(\theta,J,t+\frac{\theta-\theta_{\pm}}{\omega}\right).
\end{align}
Combining this with the fact that Eq. (\ref{eq:selfadjointcondition}) must hold at all times, we obtain the relation
\begin{equation}
 \psi\left(\theta,J,t+\frac{N\theta_b}{\omega}\right)= {\rm e}^{i N \beta\left(J\right)}\psi\left(\theta,J,t\right),
\end{equation} 
where $N$ is an integer and $\theta_b= \theta_+ -\theta_- $. When $\theta_b$ is chosen to be $\frac{2 \pi}{N}$, states may acquire an additional phase ${\rm e}^{i N \beta\left(J\right)}$ after each period of the motion. For this reason, although the dynamics are periodic, it is possible to choose boundary conditions such that the wavefunction is \emph{not} periodically symmetric due to the additional phase. While observable expectations will be unaffected by such a phase, the time translation symmetry of correlation functions will be altered by its presence. 

To demonstrate this, consider two observables $A$ and $B$ which have the two-time correlation function:
\begin{align}
C(t,t^\prime)=\int \mathrm{d}\theta \mathrm{d}J \ \psi^*\left(t\right)AB\psi(t^\prime),
\end{align} 
where phase-space arguments on the right-hand side have been omitted for brevity. If one of the time arguments is shifted by the period of motion $T=\frac{2\pi}{\omega}$, the correlation function becomes
\begin{align}
C(t+T,t^\prime)&=\int \mathrm{d}\theta \mathrm{d}J \ {\rm e}^{-i N \beta\left(J\right)}\psi^*\left(t\right)AB\psi(t^\prime) \nonumber \\
&\neq C(t,t^\prime).
\end{align} 
Thus, the phase factor determining a self-adjoint extension has a nontrivial effect on time symmetries of observable correlations. This breaking of periodicity is in sharp contrast with correlation functions on the full phase space, where continuity of the wave function forces $\beta(J)=0$ and hence $C(t,t^\prime)=C(t+T,t^\prime)$. 

The existence of a dynamical phase due to the boundary conditions is somewhat analogous to
the Berry phase \citep{Samuel1988} and its classical equivalent
the Hannay angle \citep{HANNAY1985}. The origin of these geometric
phases is quite different, resulting from adiabatic holonomic variation of the Hamiltonian parameters. In both cases however, the naive expectation that the system will return to its original state (after either a single period of the motion, or returning to the original Hamiltonian parameters) is not always true. It is well known that, for the simple harmonic oscillator, the geometric
phase change is zero \citep{0305-4470-29-10-035}, whereas here a phase may be acquired by the choice of boundary conditions. 

\subsection{Entropy conservation on the half-line \label{sec:half-line entropy}}
Angular restrictions in the action-angle coordinate system may correspond to unphysical restrictions in the momentum subspace of the phase space. A restriction purely in the coordinate of phase space may be obtained by choosing the boundaries 
\begin{equation}
J_{b}=+\infty,\qquad\theta_{-}=0,\qquad\theta_{+}=\pi,
\end{equation} 
which leads to the half-line $q\ge0$ phase space in the original coordinates. The phase space is  then ${\cal P}=\mathbb{R}_{+}\times\mathbb{R}$
and the Hilbert space is ${\cal H}=L^{2}({\cal P},{\rm d}q{\rm d}p)$. Converting the wavefunction back to its original coordinates at the boundary we obtain
\begin{equation}
\psi\left(\theta_{\pm},J\right)\to\psi\left(0,\pm p\right)
\end{equation}
which may be substituted into Eq. (\ref{eq:selfadjointcondition}) to give the domain of self-adjoint evolutions on the half-line: 
\begin{equation}
\psi(0,-p)=e^{i\beta(p)}\psi(0,p).\label{eq:wfboundarycondition}
\end{equation}
Entropy preserving probability distributions therefore
obey the condition
\begin{equation}
\rho\left(0,-p\right)=\rho\text{\ensuremath{\left(0,p\right)}}.\label{eq:reflectingcondition}
\end{equation} 
which automatically satisfies Eq. (\ref{eq:boundaries self-adjoint}).

This condition can be given a physical interpretation by considering the Liouville evolution as a continuity equation, 
where $\mathbf{j}$ is the probability current in phase space:
\begin{align}
\frac{\partial\rho}{\partial t}=-\nabla \cdot \mathbf{j},
\\ \mathbf{j}=\left(\begin{matrix} j_q \\ j_p \end{matrix}\right)=\left(\begin{matrix} \frac{\partial H}{\partial p} \\ -\frac{\partial H}{\partial q} \end{matrix}\right) \rho.
\end{align} 
The probability current in the $q$ direction (integrated over $p$) $\mathcal{J}_{q}\left[\rho\left(q,p\right)\right]$ is given by 
\begin{equation}
\mathcal{J}_{q}\left[\rho\left(q,p\right)\right]=\int{\rm d}p\ j_{q}\left(q,p\right)=\int{\rm d}p\ p\rho\left(q,p\right). \label{probcurrent}
\end{equation}
 Substituting Eq. (\ref{eq:reflectingcondition}) into the above equation,
we find that at the $q=0$ boundary, 
\begin{align}
\mathcal{J}_{q}\left[\rho\left(0,p\right)\right]=0.\label{eq:probcurrentboundary}
 \end{align}
 From this boundary condition, we conclude that the domain of states for self-adjoint Koopman operators corresponds precisely to a reflecting boundary condition at the $q=0$
boundary \citep{risken1989}.

Finally, we note that there is no $\omega$ dependence in Eqs. (\ref{eq:wfboundarycondition}) and (\ref{eq:probcurrentboundary}). In the limit $\omega\to0$, the boundary condition for entropy preserving distributions is therefore unaffected, but the system Hamiltonian now describes a free particle rather than a harmonic oscillator. For this reason, Eq. (\ref{eq:probcurrentboundary}) \emph{also} describes the self-adjoint domain of the free particle. 

\section{nonconserving States \label{sec:nonconserving states}}
Given the self-adjoint conditions derived in the previous section,
one can construct a reasonable initial
wavefunction (and hence probability density) on the half line which
does not conserve entropy. Take for example the initial wavefunction 
\begin{equation}
\psi\left(q,p\right)=Z^{-\frac{1}{2}}e^{-\frac{1}{2}(p-p_{0})^{2}-\frac{1}{2}q^{2}},
\end{equation}
where crucially, $p_{0}\ne0$ and the normalization factor is $Z=\pi/2$.
The rate of entropy change for this state can be calculated directly
from Eq. (\ref{entropyrate}). For the free particle we have in the domain $q\geq 0$
\begin{eqnarray*}
\dot{S}_{{\rm FP}} & = & -p_{0}Z^{-1}\sqrt{\pi}\left(\frac{1}{2}-\ln Z\right)\\
 & = & -p_{0}\frac{2}{\sqrt{\pi}}\ln\frac{2\sqrt{e}}{\pi}\approx-0.05p_{0}\ne0,
\end{eqnarray*}
 and for the harmonic oscillator in the same domain (setting $\omega=1$)
\begin{eqnarray*}
\dot{S}_{{\rm HO}} & = & \sqrt{\pi}p_{0}Z^{-1}\left(\ln Z+\frac{5}{2}+4p_{0}^{2}\right)\\
 & = & \frac{2p_{0}}{\sqrt{\pi}}\left(\ln\left(\frac{\pi}{2}\right)+\frac{5}{2}+4p_{0}^{2}\right).
\end{eqnarray*}
In both cases the entropy is nonconserved. The energy expectation and partition function for
both systems are also time dependent. For this state the integrated boundary probability current is
\begin{align}
\mathcal{J}_{q}\left[\rho\left(0,p\right)\right]=-2\sqrt{\pi}p_{0},
\end{align}
 i.e., at the boundary the probability is being either partially absorbed
or generated depending on the sign of $p_{0}$.

\section{Interpreting nonconserving States}
To give an interpretation to the nonconserving states, we now consider the boundary conditions for a quantum-mechanical system in the full phase space $\mathbb{R}^2$. The dynamics are governed by the Hamiltonian  $H\left(q,p\right)=H_{0}\left(q,p\right)+U_{0}\Theta\text{\ensuremath{\left(-q\right)}}$,
 where $H_0$ is the Hamiltonian for either a free-particle or harmonic oscillator,  $\Theta$ is the Heaviside step function, and $U_0$ is a constant that satisfies 
\begin{equation}
\sup_{\rho\left(q,p\right)>0}\left[H_{0}\left(q,p\right)\right]<U_{0}, \label{eq:barrierinequality}
\end{equation}
i.e., the $U_{0}\Theta\text{\ensuremath{\left(-q\right)}}$ term represents a classically impenetrable barrier on the negative part of the real line.

We describe the quantum system with the wavefunction $\psi_Q$. This can be brought into contact with earlier results by using the Wigner distribution $W\left(x,p\right)$ to describe the quantum system in phase space
\citep{Case2008}:

\begin{equation}
W\left(q,p\right)=\frac{1}{2\pi\hbar}\int{\rm d}y\ {\rm e}^{-\frac{i}{\hbar}py}\psi_{Q}\left(q+\frac{y}{2}\right)\psi_{Q}^{*}\left(q-\frac{y}{2}\right)
\end{equation}
The Wigner distribution is a quasiprobability that captures quantum-mechanical expectations
as a distribution over phase space, which in the classical limit becomes the classical wavefunction \citep{Bondar2013a}. Here, the classical wavefunction refers to the wavefunction evolved according to KvN dynamics, i.e., $\psi$ from Secs. \ref{sec:KvN dynamics}-\ref{sec:nonconserving states}.

For an arbitrary quantum operator $\hat{A}$ its expectation may be described by using the Wigner distribution:
\begin{align}
\langle \hat{A}\rangle &=\int{\rm d}q{\rm d}p\ W\left(q,p\right)A\left(q,p\right), 
\end{align}
where $A\left(q,p\right) $ is the Wigner map of the operator $\hat{A}$:
\begin{align}
A\left(q,p\right)&=\frac{1}{2\pi\hbar}\int{\rm d}y\ {\rm e}^{-\frac{i}{\hbar}py}\left\langle q+\frac{y}{2}\left|\hat{A}\right|q-\frac{y}{2}\right\rangle .
\end{align}
 The Wigner distribution evolution is described by the Moyal bracket \citep{Bondar2013a}:
\begin{align}
\frac{\partial W}{\partial t}&=\left\{ \left\{ H,W\right\} \right\} \nonumber \\ &=\frac{2}{\hbar}H\left(q,p\right)\sin\left(\frac{\hbar}{2}\overleftarrow{\frac{\partial}{\partial q}}\overrightarrow{\frac{\partial}{\partial p}}-\frac{\hbar}{2}\overleftarrow{\frac{\partial}{\partial p}}\overrightarrow{\frac{\partial}{\partial q}}\right)W\left(q,p\right).
\end{align}
When the system Hamiltonian is at most quadratic in both the $p$ and $q$ coordinates
(as is the case for the example Hamiltonians studied in Sec. IV), the Moyal bracket
is simplified, and the Wigner distribution is evolved by the Poisson bracket \citep{Case2008}, in the same way as Eq. (\ref{eq:Liouvilleevolution}):
\begin{equation}
\frac{\partial W}{\partial t}=\left\{ H,W\right\} .
\end{equation}

Remarkably, in this case, the quantum Wigner distribution, classical
probability distribution, and classical wave function \emph{all }share
the same equation of motion. In this scenario, quantum and classical
systems are distinguished purely by the restrictions placed on them
by their boundary conditions. 

\begin{figure}
\includegraphics[height=4cm]{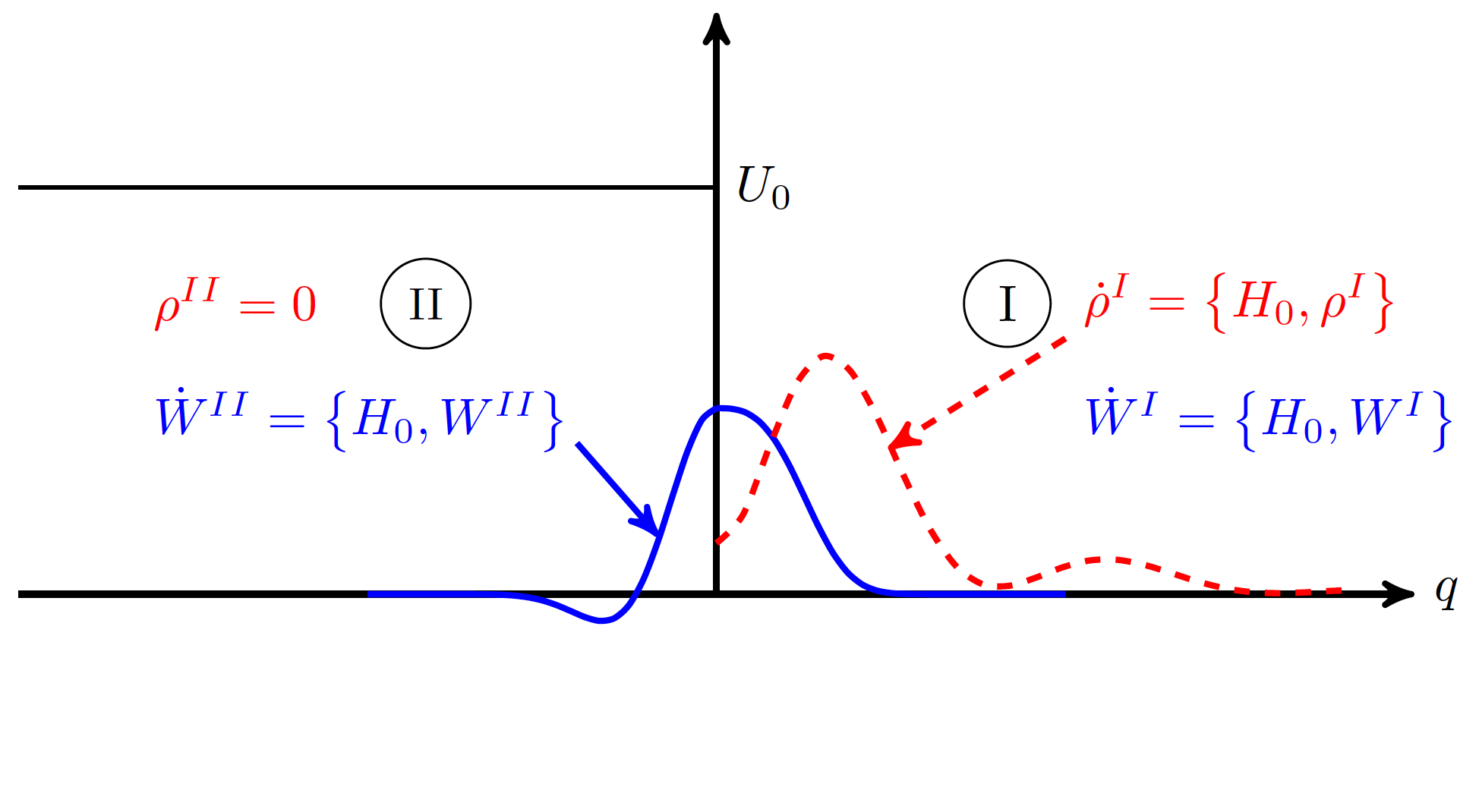} \caption{ Schematic of allowed states with an additional potential barrier $U_{0}\Theta\left(-q\right)$ in both the quantum and classical regimes. The space is partitioned into regions I and II. $\rho^I(q,p)$ is a probability distribution satisfying Eq. (\ref{eq:probcurrentboundary}), the reflecting boundary condition, and is therefore confined to region I. The same dynamics also describe a quantum Wigner distribution $W(q,p)$.  In the quantum case however, more general boundary conditions are given by Eqs. (\ref{Wignercontinuity}) and \ref{Wignercurrentboundary}), which allow for tunneling, and nonzero Wigner distributions in the classically forbidden region II. \label{fig:potential}}
\end{figure}

Labeling the positive half-line as region I and the negative half-line as region II (see Fig. \ref{fig:potential}), we consider a system initially confined to region I. By Eq. (\ref{eq:barrierinequality}), it is impossible for a classical system to transition into region II. In this case the classical system is described by the phase space  ${\cal P}=\mathbb{R}_{+}\times\mathbb{R}$ and the results of section \ref{sec:half-line entropy} apply. Most importantly, Eq. (\ref{eq:probcurrentboundary}) holds, meaning that at all times $\rho^{II}\left(q,p\right)=0$ and $\mathcal{J}_{q}\left[\rho^I\left(0,p\right)\right]=0$.

For the quantum system, while the equation of motion for $W\left(q,p\right)$ is identical to that for both the classical wave function $\psi$ and the associated probability density $\rho$, a quantum state initially confined to region I may tunnel into the classically forbidden region II. For such a quantum system, the boundary condition between regions I and II is not described by Eq. (\ref{eq:barrierinequality}), but must be generalized. This quantum boundary condition simply enforces continuity for the quantum wavefunction $\psi_Q(q)$ on the border between regions I and II \citep{Griffiths}:
\begin{align}
    \psi^I_Q (0) &= \psi^{II}_Q(0), \label{qbound1}\\
    \left.\frac{\partial \psi^I_Q (q)}{\partial q}\right|_{q=0}&= \left.\frac{\partial \psi^{II}_Q (q)}{\partial q}\right|_{q=0}. \label{qbound2}
\end{align} 
To account for this boundary condition, the Wigner distribution over the whole real-line is expressed piecewise
\begin{equation}
W\left(q,p\right)=\left\{ \begin{array}{l}
W^{I}\left(q,p\right)\textrm{ for }q\geq0\\
W^{II}\left(q,p\right)\textrm{  for }q<0
\end{array}\right. .
\end{equation} 
The quantum boundary conditions can then be expressed in terms of the Wigner function with \citep{Case2008}
\begin{equation}
    \psi_Q(q)=\frac{1}{\psi^*(0)} \int^{\infty}_{-\infty} \textrm{d}p \  \textrm{e} ^{\frac{i}{\hbar}qp} W(\frac{q}{2},p)
\end{equation}
and 
\begin{align}
   \left. \frac{\partial \psi_Q(q)}{\partial q}\right|_{q=0}=&\frac{i}{\hbar \psi^*(0)} \int^{\infty}_{-\infty} \textrm{d}p \ p W(0,p) \nonumber \\ &+\frac{1}{\psi^*(0)}\left.\frac{\partial}{\partial q}\right|_{q=0} \int^{\infty}_{-\infty} \textrm{d}p \ W(q,p).
\end{align}
Substituting these expressions into the quantum boundary conditions of Eqs. (\ref{qbound1}) and (\ref{qbound2}), yields these boundary conditions in terms of Wigner distributions:

\begin{align}
\int^{\infty}_{-\infty} \textrm{d} p \ W^{I}\left(0,p\right)=& \int^{\infty}_{-\infty} \textrm{d} p \ W^{II}\left(0,p\right), \label{Wignercontinuity} \\
    \mathcal{J}_q\left[W^{I}(0,p)-W^{II}(0,p)\right]=& \nonumber\\
    -i \hbar \left.\frac{\partial}{\partial q}\right|_{q=0} \int^{\infty}_{-\infty} \textrm{d} p \ & \left( W^{II}(q,p)-W^{I}(q,p)\right), \label{Wignercurrentboundary}
\end{align}
where the left-hand side of the second boundary condition is the Wigner flow \citep{WignerFlow, anharmonicwigner} over the boundary, defined in the same way as Eq. (\ref{eq:probcurrentboundary}). Focusing exclusively on region I, the current boundary condition is:
\begin{equation}
\mathcal{J}_q\left[W^{I}(0,p)\right]= i \hbar \left.\frac{\partial}{\partial q}\right|_{q=0} \int^{\infty}_{-\infty} \textrm{d} p \  W^{I}(q,p) +F\left[W^{II}\right]. \label{WignercurrentregionI}
\end{equation}
Here $F\left[W^{II}\right]$ is some functional dependent only on the Wigner distribution of region II.  This boundary condition is the \emph{only} feature that distinguishes $W^I$ from the classical system described by $\rho=\rho^I$. Furthermore, in the classical limit, $\hbar\to 0$ and (for a system initially confined to region I), $W^{II}=0$. In this case Eq. (\ref{WignercurrentregionI}) is equivalent to Eq. (\ref{eq:probcurrentboundary}), recovering the classical boundary condition.

Equation (\ref{WignercurrentregionI}) also allows one to interpret the entropy nonconserving states in Sec. \ref{sec:nonconserving states}. These states are characterized by $\mathcal{J}_{q}\left[\rho\left(0,p\right)\right] \neq 0$. While this violates the classical boundary condition, if $\rho$ is interpreted instead as $W^I$, then the non-zero integrated current is consistent with Eq.(\ref{WignercurrentregionI}). Hence, for the free-particle and harmonic oscillator, the entropy nonconserving states in a classical system with a classically impassible potential barrier are partial solutions for a quantum system tunneling through the barrier.

Finally, we note that the arguments presented in this section may be applied to the phenomenon of reflection above a barrier. This is another purely quantum effect, which may be regarded as tunneling through a momentum space barrier \citep{Jaffe2010}. Performing an analogous classical analysis for a harmonic system restricted to the half-line in momentum space, a similar relationship between allowed states and boundary conditions can be obtained, where classically forbidden reflected Wigner distributions match entropy nonconserving states of the classical system.

\section{Conclusions}

We have applied the Koopman von Neumann formalism to reduce the problem
of entropy conservation in a classical system to the problem of identifying self-adjoint
extensions of the Koopman operator. In this way, one can explore for a given system the full range of admissible, physical probability distributions and
phase-space restrictions  which preserve entropy.
Applying this technique to the harmonic oscillator and free particle,
a relationship between the choice of self-adjoint extension and the boundary condition was determined. In the case of the harmonic oscillator, this choice is reflected in the breaking of the periodic symmetry of correlation functions. This demonstrates a new class of situations where self-adjoint extensions of operators play an important role in distinguishing subtle phenomena. 

In the example cases studied, it was possible to construct apparently reasonable states in the phase space ${\cal P}=\mathbb{R}_{+}\times\mathbb{R}$ which do not belong to the domain of a self-adjoint Koopman operator, and consequently do not preserve entropy. These states were interpreted using the fact that the restricted phase space for the classical system corresponds to a classically impenetrable potential barrier, but a quantum system may tunnel into this forbidden region. 

For both the free particle and harmonic oscillator, quantum and classical dynamical equations coincide, and the two regimes are distinguished \emph{purely} by the allowed boundary conditions. This allowed for the interpretation of entropy nonconserving classical states as partial descriptions of tunneling quantum states. Additionally, this approach demonstrated that the classical boundary condition corresponding to a self-adjoint Koopman operator is the classical limit of the quantum boundary condition for a tunneling state.

These results highlight the importance of boundary conditions for fundamental aspects of Hamiltonian evolution. While Hamilton's equations of motion provide a local description of the dynamics, the entropy provides a global characterization of the evolution, and is therefore sensitive to the self-adjointness of the Koopman operator. Even in the case that equations of motion are formally time-reversal symmetric, the choice of boundary condition can break entropy preservation, and thus time-reversal symmetry.

\begin{acknowledgments}
The authors would like to thank the referees for their valuable comments, which have inspired a number of the results presented here. D.I.B. is grateful to Jean-Luc Cambier for suggesting a research direction that has directly led to this work. G.M. and D.I.B. are supported by Air Force Office of Scientific Research Young Investigator Research Program (FA9550-16-1-0254). A.P. is supported by project No. 1.669.2016/1.4 of the Ministry of Science and Higher Education of the Russian Federation. 
\end{acknowledgments}

\bibliographystyle{aipnum4-1}

\begin{thebibliography}{0}%
\makeatletter
\providecommand \@ifxundefined [1]{%
 \@ifx{#1\undefined}
}%
\providecommand \@ifnum [1]{%
 \ifnum #1\expandafter \@firstoftwo
 \else \expandafter \@secondoftwo
 \fi
}%
\providecommand \@ifx [1]{%
 \ifx #1\expandafter \@firstoftwo
 \else \expandafter \@secondoftwo
 \fi
}%
\providecommand \natexlab [1]{#1}%
\providecommand \enquote  [1]{``#1''}%
\providecommand \bibnamefont  [1]{#1}%
\providecommand \bibfnamefont [1]{#1}%
\providecommand \citenamefont [1]{#1}%
\providecommand \href@noop [0]{\@secondoftwo}%
\providecommand \href [0]{\begingroup \@sanitize@url \@href}%
\providecommand \@href[1]{\@@startlink{#1}\@@href}%
\providecommand \@@href[1]{\endgroup#1\@@endlink}%
\providecommand \@sanitize@url [0]{\catcode `\\12\catcode `\$12\catcode
  `\&12\catcode `\#12\catcode `\^12\catcode `\_12\catcode `\%12\relax}%
\providecommand \@@startlink[1]{}%
\providecommand \@@endlink[0]{}%
\providecommand \url  [0]{\begingroup\@sanitize@url \@url }%
\providecommand \@url [1]{\endgroup\@href {#1}{\urlprefix }}%
\providecommand \urlprefix  [0]{URL }%
\providecommand \Eprint [0]{\href }%
\providecommand \doibase [0]{http://dx.doi.org/}%
\providecommand \selectlanguage [0]{\@gobble}%
\providecommand \bibinfo  [0]{\@secondoftwo}%
\providecommand \bibfield  [0]{\@secondoftwo}%
\providecommand \translation [1]{[#1]}%
\providecommand \BibitemOpen [0]{}%
\providecommand \bibitemStop [0]{}%
\providecommand \bibitemNoStop [0]{.\EOS\space}%
\providecommand \EOS [0]{\spacefactor3000\relax}%
\providecommand \BibitemShut  [1]{\csname bibitem#1\endcsname}%
\let\auto@bib@innerbib\@empty
\end{thebibliography}%


\begin{thebibliography}{58}%
	\makeatletter
	\providecommand \@ifxundefined [1]{%
		\@ifx{#1\undefined}
	}%
	\providecommand \@ifnum [1]{%
		\ifnum #1\expandafter \@firstoftwo
		\else \expandafter \@secondoftwo
		\fi
	}%
	\providecommand \@ifx [1]{%
		\ifx #1\expandafter \@firstoftwo
		\else \expandafter \@secondoftwo
		\fi
	}%
	\providecommand \natexlab [1]{#1}%
	\providecommand \enquote  [1]{``#1''}%
	\providecommand \bibnamefont  [1]{#1}%
	\providecommand \bibfnamefont [1]{#1}%
	\providecommand \citenamefont [1]{#1}%
	\providecommand \href@noop [0]{\@secondoftwo}%
	\providecommand \href [0]{\begingroup \@sanitize@url \@href}%
	\providecommand \@href[1]{\@@startlink{#1}\@@href}%
	\providecommand \@@href[1]{\endgroup#1\@@endlink}%
	\providecommand \@sanitize@url [0]{\catcode `\\12\catcode `\$12\catcode
		`\&12\catcode `\#12\catcode `\^12\catcode `\_12\catcode `\%12\relax}%
	\providecommand \@@startlink[1]{}%
	\providecommand \@@endlink[0]{}%
	\providecommand \url  [0]{\begingroup\@sanitize@url \@url }%
	\providecommand \@url [1]{\endgroup\@href {#1}{\urlprefix }}%
	\providecommand \urlprefix  [0]{URL }%
	\providecommand \Eprint [0]{\href }%
	\providecommand \doibase [0]{http://dx.doi.org/}%
	\providecommand \selectlanguage [0]{\@gobble}%
	\providecommand \bibinfo  [0]{\@secondoftwo}%
	\providecommand \bibfield  [0]{\@secondoftwo}%
	\providecommand \translation [1]{[#1]}%
	\providecommand \BibitemOpen [0]{}%
	\providecommand \bibitemStop [0]{}%
	\providecommand \bibitemNoStop [0]{.\EOS\space}%
	\providecommand \EOS [0]{\spacefactor3000\relax}%
	\providecommand \BibitemShut  [1]{\csname bibitem#1\endcsname}%
	\let\auto@bib@innerbib\@empty
	\bibitem [{\citenamefont {Lieb}\ and\ \citenamefont
		{Yngvason}(2000)}]{Lieb2000}%
	\BibitemOpen
	\bibfield  {author} {\bibinfo {author} {\bibfnamefont {E.~H.}\ \bibnamefont
			{Lieb}}\ and\ \bibinfo {author} {\bibfnamefont {J.}~\bibnamefont
			{Yngvason}},\ }\href {\doibase 10.1063/1.883034} {\bibfield  {journal}
		{\bibinfo  {journal} {Phys. Today}\ }\textbf {\bibinfo {volume} {53}},\
		\bibinfo {pages} {32} (\bibinfo {year} {2000})}\BibitemShut {NoStop}%
	\bibitem [{\citenamefont {Ford}(2015)}]{Ford2015}%
	\BibitemOpen
	\bibfield  {author} {\bibinfo {author} {\bibfnamefont {I.~J.}\ \bibnamefont
			{Ford}},\ }\href {\doibase 10.1088/1367-2630/17/7/075017} {\bibfield
		{journal} {\bibinfo  {journal} {New J. Phys.}\ }\textbf {\bibinfo {volume}
			{17}},\ \bibinfo {pages} {075017} (\bibinfo {year} {2015})}\BibitemShut
	{NoStop}%
	\bibitem [{\citenamefont {Parrondo}, \citenamefont {den Broeck},\ and\
		\citenamefont {Kawai}(2009)}]{1367-2630-11-7-073008}%
	\BibitemOpen
	\bibfield  {author} {\bibinfo {author} {\bibfnamefont {J.~M.~R.}\
			\bibnamefont {Parrondo}}, \bibinfo {author} {\bibfnamefont {C.~V.}\
			\bibnamefont {den Broeck}}, \ and\ \bibinfo {author} {\bibfnamefont
			{R.}~\bibnamefont {Kawai}},\ }\href
	{http://stacks.iop.org/1367-2630/11/i=7/a=073008} {\bibfield  {journal}
		{\bibinfo  {journal} {New J. Phys.}\ }\textbf {\bibinfo {volume} {11}},\
		\bibinfo {pages} {73008} (\bibinfo {year} {2009})}\BibitemShut {NoStop}%
	\bibitem [{\citenamefont {{Brown}}, \citenamefont {{Myrvold}},\ and\
		\citenamefont {{Uffink}}(2009)}]{Timearrow}%
	\BibitemOpen
	\bibfield  {author} {\bibinfo {author} {\bibfnamefont {H.~R.}\ \bibnamefont
			{{Brown}}}, \bibinfo {author} {\bibfnamefont {W.}~\bibnamefont {{Myrvold}}},
		\ and\ \bibinfo {author} {\bibfnamefont {J.}~\bibnamefont {{Uffink}}},\
	}\href {\doibase 10.1016/j.shpsb.2009.03.003} {\bibfield  {journal} {\bibinfo
			{journal} {Stud. Hist. Philos. Sci.}\ }\textbf {\bibinfo {volume} {40}},\
		\bibinfo {pages} {174} (\bibinfo {year} {2009})},\ \Eprint
	{http://arxiv.org/abs/0809.1304} {arXiv:0809.1304} \BibitemShut {NoStop}%
	\bibitem [{\citenamefont {Batalh{\~{a}}o}\ \emph {et~al.}(2015)\citenamefont
		{Batalh{\~{a}}o}, \citenamefont {Souza}, \citenamefont {Sarthour},
		\citenamefont {Oliveira}, \citenamefont {Paternostro}, \citenamefont {Lutz},\
		and\ \citenamefont {Serra}}]{Batalhao2015}%
	\BibitemOpen
	\bibfield  {author} {\bibinfo {author} {\bibfnamefont {T.~B.}\ \bibnamefont
			{Batalh{\~{a}}o}}, \bibinfo {author} {\bibfnamefont {A.~M.}\ \bibnamefont
			{Souza}}, \bibinfo {author} {\bibfnamefont {R.~S.}\ \bibnamefont {Sarthour}},
		\bibinfo {author} {\bibfnamefont {I.~S.}\ \bibnamefont {Oliveira}}, \bibinfo
		{author} {\bibfnamefont {M.}~\bibnamefont {Paternostro}}, \bibinfo {author}
		{\bibfnamefont {E.}~\bibnamefont {Lutz}}, \ and\ \bibinfo {author}
		{\bibfnamefont {R.~M.}\ \bibnamefont {Serra}},\ }\href {\doibase
		10.1103/PhysRevLett.115.190601} {\bibfield  {journal} {\bibinfo  {journal}
			{Phys. Rev Lett.}\ }\textbf {\bibinfo {volume} {115}},\ \bibinfo {pages}
		{190601} (\bibinfo {year} {2015})},\ \Eprint
	{http://arxiv.org/abs/1502.06704} {arXiv:1502.06704} \BibitemShut {NoStop}%
	\bibitem [{\citenamefont {Hanggi}, \citenamefont {Hilbert},\ and\ \citenamefont
		{Dunkel}(2015)}]{Hanggi}%
	\BibitemOpen
	\bibfield  {author} {\bibinfo {author} {\bibfnamefont {P.}~\bibnamefont
			{Hanggi}}, \bibinfo {author} {\bibfnamefont {S.}~\bibnamefont {Hilbert}}, \
		and\ \bibinfo {author} {\bibfnamefont {J.}~\bibnamefont {Dunkel}},\
	}\href@noop {} {\bibfield  {journal} {\bibinfo  {journal} {Philos. Trans.
				Royal Soc. A}\ }\textbf {\bibinfo {volume} {374}} (\bibinfo {year}
		{2015})}\BibitemShut {NoStop}%
	\bibitem [{\citenamefont {Jaynes}(1992)}]{Jaynes1992}%
	\BibitemOpen
	\bibfield  {author} {\bibinfo {author} {\bibfnamefont {E.~E.~T.}\
			\bibnamefont {Jaynes}},\ }in\ \href {\doibase 10.1007/BF00254285} {\emph
		{\bibinfo {booktitle} {Maximum entropy and Bayesian methods}}},\
	Vol.~\bibinfo {volume} {17}\ (\bibinfo  {publisher} {Springer},\ \bibinfo
	{year} {1992})\ pp.\ \bibinfo {pages} {1--21}\BibitemShut {NoStop}%
	\bibitem [{\citenamefont {Mackey}(1989)}]{RevModPhys.61.981}%
	\BibitemOpen
	\bibfield  {author} {\bibinfo {author} {\bibfnamefont {M.~C.}\ \bibnamefont
			{Mackey}},\ }\href {\doibase 10.1103/RevModPhys.61.981} {\bibfield  {journal}
		{\bibinfo  {journal} {Rev. Mod. Phys.}\ }\textbf {\bibinfo {volume} {61}},\
		\bibinfo {pages} {981} (\bibinfo {year} {1989})}\BibitemShut {NoStop}%
	\bibitem [{\citenamefont {Landau}\ and\ \citenamefont
		{Lifshitz}(1951)}]{LandauLifshitz}%
	\BibitemOpen
	\bibfield  {author} {\bibinfo {author} {\bibfnamefont {L.}~\bibnamefont
			{Landau}}\ and\ \bibinfo {author} {\bibfnamefont {E.}~\bibnamefont
			{Lifshitz}},\ }\href@noop {} {\emph {\bibinfo {title} {Course of Theoretical
				Physics: Statistical Physics}}},\ Vol.~\bibinfo {volume} {5}\ (\bibinfo
	{publisher} {Elsevier},\ \bibinfo {address} {London},\ \bibinfo {year}
	{1951})\BibitemShut {NoStop}%
	\bibitem [{\citenamefont {Bonneau}, \citenamefont {Faraut},\ and\ \citenamefont
		{Valent}(2001)}]{Bonneau2001}%
	\BibitemOpen
	\bibfield  {author} {\bibinfo {author} {\bibfnamefont {G.}~\bibnamefont
			{Bonneau}}, \bibinfo {author} {\bibfnamefont {J.}~\bibnamefont {Faraut}}, \
		and\ \bibinfo {author} {\bibfnamefont {G.}~\bibnamefont {Valent}},\ }\href
	{\doibase 10.1119/1.1328351} {\bibfield  {journal} {\bibinfo  {journal} {Am.
				J. Phys.}\ }\textbf {\bibinfo {volume} {69}},\ \bibinfo {pages} {322}
		(\bibinfo {year} {2001})},\ \Eprint {http://arxiv.org/abs/0103153}
	{arXiv:0103153} \BibitemShut {NoStop}%
	\bibitem [{\citenamefont {Chen}, \citenamefont {Lu},\ and\ \citenamefont
		{Vishwanath}(2014)}]{Chen2014}%
	\BibitemOpen
	\bibfield  {author} {\bibinfo {author} {\bibfnamefont {X.}~\bibnamefont
			{Chen}}, \bibinfo {author} {\bibfnamefont {Y.~M.}\ \bibnamefont {Lu}}, \ and\
		\bibinfo {author} {\bibfnamefont {A.}~\bibnamefont {Vishwanath}},\ }\href
	{\doibase 10.1038/ncomms4507} {\bibfield  {journal} {\bibinfo  {journal}
			{Nat. Commun.}\ }\textbf {\bibinfo {volume} {5}},\ \bibinfo {pages} {1}
		(\bibinfo {year} {2014})},\ \Eprint {http://arxiv.org/abs/1303.4301}
	{arXiv:1303.4301} \BibitemShut {NoStop}%
	\bibitem [{\citenamefont {Chen}\ \emph {et~al.}(2012)\citenamefont {Chen},
		\citenamefont {Gu}, \citenamefont {Liu},\ and\ \citenamefont
		{Wen}}]{Chen2012}%
	\BibitemOpen
	\bibfield  {author} {\bibinfo {author} {\bibfnamefont {X.}~\bibnamefont
			{Chen}}, \bibinfo {author} {\bibfnamefont {Z.~C.}\ \bibnamefont {Gu}},
		\bibinfo {author} {\bibfnamefont {Z.~X.}\ \bibnamefont {Liu}}, \ and\
		\bibinfo {author} {\bibfnamefont {X.~G.}\ \bibnamefont {Wen}},\ }\href
	{\doibase 10.1126/science.1227224} {\bibfield  {journal} {\bibinfo  {journal}
			{Science}\ }\textbf {\bibinfo {volume} {338}},\ \bibinfo {pages} {1604}
		(\bibinfo {year} {2012})},\ \Eprint {http://arxiv.org/abs/1301.0861}
	{arXiv:1301.0861} \BibitemShut {NoStop}%
	\bibitem [{\citenamefont {Ahari}, \citenamefont {Ortiz},\ and\ \citenamefont
		{Seradjeh}(2016)}]{Ahari2016}%
	\BibitemOpen
	\bibfield  {author} {\bibinfo {author} {\bibfnamefont {M.~T.}\ \bibnamefont
			{Ahari}}, \bibinfo {author} {\bibfnamefont {G.}~\bibnamefont {Ortiz}}, \ and\
		\bibinfo {author} {\bibfnamefont {B.}~\bibnamefont {Seradjeh}},\ }\href
	{\doibase 10.1119/1.4961500} {\bibfield  {journal} {\bibinfo  {journal} {Am.
				J. Phys.}\ }\textbf {\bibinfo {volume} {84}},\ \bibinfo {pages} {858}
		(\bibinfo {year} {2016})},\ \Eprint {http://arxiv.org/abs/1508.02682}
	{arXiv:1508.02682} \BibitemShut {NoStop}%
	\bibitem [{\citenamefont {Araujo}, \citenamefont {Coutinho},\ and\
		\citenamefont {{Fernando Perez}}(2004)}]{Araujo2004}%
	\BibitemOpen
	\bibfield  {author} {\bibinfo {author} {\bibfnamefont {V.~S.}\ \bibnamefont
			{Araujo}}, \bibinfo {author} {\bibfnamefont {F.~A.~B.}\ \bibnamefont
			{Coutinho}}, \ and\ \bibinfo {author} {\bibfnamefont {J.}~\bibnamefont
			{{Fernando Perez}}},\ }\href {\doibase 10.1119/1.1624111} {\bibfield
		{journal} {\bibinfo  {journal} {Am. J. Phys.}\ }\textbf {\bibinfo {volume}
			{72}},\ \bibinfo {pages} {203} (\bibinfo {year} {2004})}\BibitemShut
	{NoStop}%
	\bibitem [{\citenamefont {Berman}(1991)}]{Berman1991}%
	\BibitemOpen
	\bibfield  {author} {\bibinfo {author} {\bibfnamefont {D.~H.}\ \bibnamefont
			{Berman}},\ }\href@noop {} {\bibfield  {journal} {\bibinfo  {journal} {Am. J.
				Phys.}\ }\textbf {\bibinfo {volume} {59}},\ \bibinfo {pages} {937} (\bibinfo
		{year} {1991})}\BibitemShut {NoStop}%
	\bibitem [{\citenamefont {Capri}(1977)}]{Capri1977}%
	\BibitemOpen
	\bibfield  {author} {\bibinfo {author} {\bibfnamefont {A.~Z.}\ \bibnamefont
			{Capri}},\ }\href {\doibase 10.1119/1.11055} {\bibfield  {journal} {\bibinfo
			{journal} {Am. J. Phys.}\ }\textbf {\bibinfo {volume} {45}},\ \bibinfo
		{pages} {823} (\bibinfo {year} {1977})}\BibitemShut {NoStop}%
	\bibitem [{\citenamefont {Bender}, \citenamefont {Brody},\ and\ \citenamefont
		{M{\"{u}}ller}(2017)}]{Bender2017}%
	\BibitemOpen
	\bibfield  {author} {\bibinfo {author} {\bibfnamefont {C.~M.}\ \bibnamefont
			{Bender}}, \bibinfo {author} {\bibfnamefont {D.~C.}\ \bibnamefont {Brody}}, \
		and\ \bibinfo {author} {\bibfnamefont {M.~P.}\ \bibnamefont {M{\"{u}}ller}},\
	}\href {\doibase 10.1103/PhysRevLett.118.130201} {\bibfield  {journal}
		{\bibinfo  {journal} {Phys. Rev Lett.}\ }\textbf {\bibinfo {volume} {118}},\
		\bibinfo {pages} {130201} (\bibinfo {year} {2017})},\ \Eprint
	{http://arxiv.org/abs/1608.03679} {arXiv:1608.03679} \BibitemShut {NoStop}%
	\bibitem [{\citenamefont {Koopman}(1931)}]{Koopman315}%
	\BibitemOpen
	\bibfield  {author} {\bibinfo {author} {\bibfnamefont {B.~O.}\ \bibnamefont
			{Koopman}},\ }\href {\doibase 10.1073/pnas.17.5.315} {\bibfield  {journal}
		{\bibinfo  {journal} {Proc. Natl. Acad. Sci. U.S.A.}\ }\textbf {\bibinfo
			{volume} {17}},\ \bibinfo {pages} {315} (\bibinfo {year} {1931})}\BibitemShut
	{NoStop}%
	\bibitem [{\citenamefont {Reed}\ and\ \citenamefont
		{Simon}(1980)}]{ReedSimon2}%
	\BibitemOpen
	\bibfield  {author} {\bibinfo {author} {\bibfnamefont {M.}~\bibnamefont
			{Reed}}\ and\ \bibinfo {author} {\bibfnamefont {B.}~\bibnamefont {Simon}},\
	}\href@noop {} {\emph {\bibinfo {title} {Methods of Modern Mathematical
				Physics}}},\ Vol.\ \bibinfo {volume} {1,2}\ (\bibinfo  {publisher}
	{Elsevier},\ \bibinfo {address} {London},\ \bibinfo {year} {1980})\ Chap.\
	\bibinfo {chapter} {2,9}\BibitemShut {NoStop}%
	\bibitem [{\citenamefont {Sudarshan}(1976)}]{Sudarshan1976}%
	\BibitemOpen
	\bibfield  {author} {\bibinfo {author} {\bibfnamefont {E.~C.~G.}\
			\bibnamefont {Sudarshan}},\ }\href {\doibase 10.1007/BF02847120} {\bibfield
		{journal} {\bibinfo  {journal} {Pramana}\ }\textbf {\bibinfo {volume} {6}},\
		\bibinfo {pages} {117} (\bibinfo {year} {1976})}\BibitemShut {NoStop}%
	\bibitem [{\citenamefont {Viennot}\ and\ \citenamefont
		{Aubourg}(2018)}]{Viennot_2018}%
	\BibitemOpen
	\bibfield  {author} {\bibinfo {author} {\bibfnamefont {D.}~\bibnamefont
			{Viennot}}\ and\ \bibinfo {author} {\bibfnamefont {L.}~\bibnamefont
			{Aubourg}},\ }\href {\doibase 10.1088/1751-8121/aaca45} {\bibfield  {journal}
		{\bibinfo  {journal} {J. Phys. A}\ }\textbf {\bibinfo {volume} {51}},\
		\bibinfo {pages} {335201} (\bibinfo {year} {2018})}\BibitemShut {NoStop}%
	\bibitem [{\citenamefont {Bondar}\ and\ \citenamefont
		{Tronci}(2018)}]{Bondarquantumclassical}%
	\BibitemOpen
	\bibfield  {author} {\bibinfo {author} {\bibfnamefont {D.~I.}\ \bibnamefont
			{Bondar}},\ \bibinfo {author} {\bibfnamefont {F.}~\bibnamefont
			{Gay-Balmaz}},\ and\ \bibinfo {author} {\bibfnamefont {C.}~\bibnamefont
			{Tronci}},\ }\href@noop {} {\  (\bibinfo {year} {2018})},\ \Eprint
	{http://arxiv.org/abs/1802.04787v2} {arXiv:1802.04787v2} \BibitemShut
	{NoStop}%
	\bibitem [{\citenamefont {Okuyama}\ and\ \citenamefont
		{Ohzeki}(2018)}]{PhysRevLett.120.070402}%
	\BibitemOpen
	\bibfield  {author} {\bibinfo {author} {\bibfnamefont {M.}~\bibnamefont
			{Okuyama}}\ and\ \bibinfo {author} {\bibfnamefont {M.}~\bibnamefont
			{Ohzeki}},\ }\href {\doibase 10.1103/PhysRevLett.120.070402} {\bibfield
		{journal} {\bibinfo  {journal} {Phys. Rev. Lett.}\ }\textbf {\bibinfo
			{volume} {120}},\ \bibinfo {pages} {070402} (\bibinfo {year}
		{2018})}\BibitemShut {NoStop}%
	\bibitem [{\citenamefont {Rajagopal}\ and\ \citenamefont
		{Ghose}(2016)}]{Hilbertelectrodynamics}%
	\BibitemOpen
	\bibfield  {author} {\bibinfo {author} {\bibfnamefont {A.~K.}\ \bibnamefont
			{Rajagopal}}\ and\ \bibinfo {author} {\bibfnamefont {P.}~\bibnamefont
			{Ghose}},\ }\href {\doibase 10.1007/s12043-015-1172-8} {\bibfield  {journal}
		{\bibinfo  {journal} {Pramana}\ }\textbf {\bibinfo {volume} {86}},\ \bibinfo
		{pages} {1161} (\bibinfo {year} {2016})}\BibitemShut {NoStop}%
	\bibitem [{\citenamefont {Bondar}\ \emph {et~al.}(2012)\citenamefont {Bondar},
		\citenamefont {Cabrera}, \citenamefont {Lompay}, \citenamefont {Ivanov},\
		and\ \citenamefont {Rabitz}}]{Bondar2012a}%
	\BibitemOpen
	\bibfield  {author} {\bibinfo {author} {\bibfnamefont {D.~I.}\ \bibnamefont
			{Bondar}}, \bibinfo {author} {\bibfnamefont {R.}~\bibnamefont {Cabrera}},
		\bibinfo {author} {\bibfnamefont {R.~R.}\ \bibnamefont {Lompay}}, \bibinfo
		{author} {\bibfnamefont {M.~Y.}\ \bibnamefont {Ivanov}}, \ and\ \bibinfo
		{author} {\bibfnamefont {H.~A.}\ \bibnamefont {Rabitz}},\ }\href {\doibase
		10.1103/PhysRevLett.109.190403} {\bibfield  {journal} {\bibinfo  {journal}
			{Phys. Rev Lett.}\ }\textbf {\bibinfo {volume} {109}},\ \bibinfo {pages}
		{190403} (\bibinfo {year} {2012})},\ \Eprint
	{http://arxiv.org/abs/1105.4014v4} {arXiv:1105.4014v4} \BibitemShut {NoStop}%
	\bibitem [{\citenamefont {Bondar}, \citenamefont {Cabrera},\ and\ \citenamefont
		{Rabitz}(2013)}]{Bondar2013}%
	\BibitemOpen
	\bibfield  {author} {\bibinfo {author} {\bibfnamefont {D.~I.}\ \bibnamefont
			{Bondar}}, \bibinfo {author} {\bibfnamefont {R.}~\bibnamefont {Cabrera}}, \
		and\ \bibinfo {author} {\bibfnamefont {H.~A.}\ \bibnamefont {Rabitz}},\
	}\href {\doibase 10.1103/PhysRevA.88.012116} {\bibfield  {journal} {\bibinfo
			{journal} {Phys. Rev. A}\ }\textbf {\bibinfo {volume} {88}},\ \bibinfo
		{pages} {012116} (\bibinfo {year} {2013})},\ \Eprint
	{http://arxiv.org/abs/1112.3679} {arXiv:1112.3679} \BibitemShut {NoStop}%
	\bibitem [{\citenamefont {Gozzi}, \citenamefont {Cattaruzza},\ and\
		\citenamefont {Pagani}(2014)}]{Gozzi2014}%
	\BibitemOpen
	\bibfield  {author} {\bibinfo {author} {\bibfnamefont {E.}~\bibnamefont
			{Gozzi}}, \bibinfo {author} {\bibfnamefont {E.}~\bibnamefont {Cattaruzza}}, \
		and\ \bibinfo {author} {\bibfnamefont {C.}~\bibnamefont {Pagani}},\ }\href
	{\doibase 10.1142/9183} {\emph {\bibinfo {title} {Path Integrals for
				Pedestrians}}}\ (\bibinfo  {publisher} {World Scientific},\ \bibinfo {year}
	{2014})\BibitemShut {NoStop}%
	\bibitem [{\citenamefont {Shee}(2015)}]{1505.06391}%
	\BibitemOpen
	\bibfield  {author} {\bibinfo {author} {\bibfnamefont {J.}~\bibnamefont
			{Shee}},\ }\href@noop {} {} (\bibinfo {year} {2015}),\ \Eprint
	{http://arxiv.org/abs/arXiv:1505.06391} {arXiv:1505.06391} \BibitemShut
	{NoStop}%
	\bibitem [{\citenamefont {Gozzi}\ and\ \citenamefont
		{Reuter}(1994)}]{Gozzi1994}%
	\BibitemOpen
	\bibfield  {author} {\bibinfo {author} {\bibfnamefont {E.}~\bibnamefont
			{Gozzi}}\ and\ \bibinfo {author} {\bibfnamefont {M.}~\bibnamefont {Reuter}},\
	}\href {\doibase 10.1016/0960-0779(94)90026-4} {\bibfield  {journal}
		{\bibinfo  {journal} {Chaos Solitons Fractals}\ }\textbf {\bibinfo {volume}
			{4}},\ \bibinfo {pages} {1117} (\bibinfo {year} {1994})}\BibitemShut
	{NoStop}%
	\bibitem [{\citenamefont {Liboff}(2003)}]{Liboff}%
	\BibitemOpen
	\bibfield  {author} {\bibinfo {author} {\bibfnamefont {R.}~\bibnamefont
			{Liboff}},\ }\href {\doibase 10.1007/b97467} {\emph {\bibinfo {title}
			{Kinetic Theory}}}\ (\bibinfo  {publisher} {Springer-Verlag},\ \bibinfo
	{year} {2003})\ Chap.~\bibinfo {chapter} {2}\BibitemShut {NoStop}%
	\bibitem [{\citenamefont {McCaul}(2018)}]{Mythesis}%
	\BibitemOpen
	\bibfield  {author} {\bibinfo {author} {\bibfnamefont {G.}~\bibnamefont
			{McCaul}},\ }\emph {\bibinfo {title} {Stochastic Representations of Open
			Systems}},\ \href@noop {} {Ph.D. thesis},\ \bibinfo  {school} {King's College
		London} (\bibinfo {year} {2018})\BibitemShut {NoStop}%
	\bibitem [{\citenamefont {Chru{\'{s}}ci{\'{n}}ski}(2006)}]{Chruscinski2006}%
	\BibitemOpen
	\bibfield  {author} {\bibinfo {author} {\bibfnamefont {D.}~\bibnamefont
			{Chru{\'{s}}ci{\'{n}}ski}},\ }\href {\doibase 10.1016/S0034-4877(06)80023-6}
	{\bibfield  {journal} {\bibinfo  {journal} {Rep. Math. Phys.}\ }\textbf
		{\bibinfo {volume} {57}},\ \bibinfo {pages} {319} (\bibinfo {year}
		{2006})}\BibitemShut {NoStop}%
	\bibitem [{\citenamefont {Brunton}\ \emph {et~al.}(2016)\citenamefont
		{Brunton}, \citenamefont {Brunton}, \citenamefont {Proctor},\ and\
		\citenamefont {Kutz}}]{Koopman-nonlinear}%
	\BibitemOpen
	\bibfield  {author} {\bibinfo {author} {\bibfnamefont {S.~L.}\ \bibnamefont
			{Brunton}}, \bibinfo {author} {\bibfnamefont {B.~W.}\ \bibnamefont
			{Brunton}}, \bibinfo {author} {\bibfnamefont {J.~L.}\ \bibnamefont
			{Proctor}}, \ and\ \bibinfo {author} {\bibfnamefont {J.~N.}\ \bibnamefont
			{Kutz}},\ }\href {\doibase 10.1371/journal.pone.0150171} {\bibfield
		{journal} {\bibinfo  {journal} {PLOS ONE}\ }\textbf {\bibinfo {volume}
			{11}},\ \bibinfo {pages} {e0150171} (\bibinfo {year} {2016})}\BibitemShut
	{NoStop}%
	\bibitem [{\citenamefont {Ramos-Prieto}\ \emph {et~al.}(2018)\citenamefont
		{Ramos-Prieto}, \citenamefont {Urz{\'{u}}a-Pineda}, \citenamefont
		{Soto-Eguibar},\ and\ \citenamefont {Moya-Cessa}}]{RamosPrieto2018}%
	\BibitemOpen
	\bibfield  {author} {\bibinfo {author} {\bibfnamefont {I.}~\bibnamefont
			{Ramos-Prieto}}, \bibinfo {author} {\bibfnamefont {A.~R.}\ \bibnamefont
			{Urz{\'{u}}a-Pineda}}, \bibinfo {author} {\bibfnamefont {F.}~\bibnamefont
			{Soto-Eguibar}}, \ and\ \bibinfo {author} {\bibfnamefont {H.~M.}\
			\bibnamefont {Moya-Cessa}},\ }\href@noop {} {\bibfield  {journal} {\bibinfo
			{journal} {Sci. Rep.}\ }\textbf {\bibinfo {volume} {8}},\ \bibinfo {pages}
		{8401} (\bibinfo {year} {2018})}\BibitemShut {NoStop}%
	\bibitem [{\citenamefont {Mezi{\'c}}(2005)}]{mezic2005spectral}%
	\BibitemOpen
	\bibfield  {author} {\bibinfo {author} {\bibfnamefont {I.}~\bibnamefont
			{Mezi{\'c}}},\ }\href@noop {} {\bibfield  {journal} {\bibinfo  {journal}
			{Nonlinear Dyn.}\ }\textbf {\bibinfo {volume} {41}},\ \bibinfo {pages} {309}
		(\bibinfo {year} {2005})}\BibitemShut {NoStop}%
	\bibitem [{\citenamefont {Budi{\v{s}}i{\'c}}, \citenamefont {Mohr},\ and\
		\citenamefont {Mezi{\'c}}(2012)}]{budivsic2012applied}%
	\BibitemOpen
	\bibfield  {author} {\bibinfo {author} {\bibfnamefont {M.}~\bibnamefont
			{Budi{\v{s}}i{\'c}}}, \bibinfo {author} {\bibfnamefont {R.}~\bibnamefont
			{Mohr}}, \ and\ \bibinfo {author} {\bibfnamefont {I.}~\bibnamefont
			{Mezi{\'c}}},\ }\href@noop {} {\bibfield  {journal} {\bibinfo  {journal}
			{Chaos}\ }\textbf {\bibinfo {volume} {22}},\ \bibinfo {pages} {047510}
		(\bibinfo {year} {2012})}\BibitemShut {NoStop}%
	\bibitem [{\citenamefont {Brumer}\ and\ \citenamefont
		{Gong}(2006)}]{BrumerBornRule}%
	\BibitemOpen
	\bibfield  {author} {\bibinfo {author} {\bibfnamefont {P.}~\bibnamefont
			{Brumer}}\ and\ \bibinfo {author} {\bibfnamefont {J.}~\bibnamefont {Gong}},\
	}\href {\doibase 10.1103/PhysRevA.73.052109} {\bibfield  {journal} {\bibinfo
			{journal} {Phys. Rev. A}\ }\textbf {\bibinfo {volume} {73}},\ \bibinfo
		{pages} {052109} (\bibinfo {year} {2006})}\BibitemShut {NoStop}%
	\bibitem [{\citenamefont {von Neumann}(1932{\natexlab{a}})}]{Stonetheorem}%
	\BibitemOpen
	\bibfield  {author} {\bibinfo {author} {\bibfnamefont {J.}~\bibnamefont {von
				Neumann}},\ }\href {http://www.jstor.org/stable/1968535} {\bibfield
		{journal} {\bibinfo  {journal} {Ann. Math.}\ }\textbf {\bibinfo {volume}
			{33}},\ \bibinfo {pages} {567} (\bibinfo {year}
		{1932}{\natexlab{a}})}\BibitemShut {NoStop}%
	\bibitem [{\citenamefont {Hamhalter}(2003)}]{measure}%
	\BibitemOpen
	\bibfield  {author} {\bibinfo {author} {\bibfnamefont {J.}~\bibnamefont
			{Hamhalter}},\ }\href@noop {} {\emph {\bibinfo {title} {Quantum Measure
				theory}}}\ (\bibinfo  {publisher} {Springer},\ \bibinfo {address} {New
		York},\ \bibinfo {year} {2003})\BibitemShut {NoStop}%
	\bibitem [{\citenamefont {Baker}(1958)}]{Baker1958}%
	\BibitemOpen
	\bibfield  {author} {\bibinfo {author} {\bibfnamefont {G.}~\bibnamefont
			{Baker}},\ }\href {\doibase 10.1103/PhysRev.109.2198} {\bibfield  {journal}
		{\bibinfo  {journal} {Phys. Rev.}\ }\textbf {\bibinfo {volume} {109}},\
		\bibinfo {pages} {2198} (\bibinfo {year} {1958})}\BibitemShut {NoStop}%
	\bibitem [{\citenamefont {Curtright}, \citenamefont {Fairlie},\ and\
		\citenamefont {Zachos}(2014)}]{Curtright2014}%
	\BibitemOpen
	\bibfield  {author} {\bibinfo {author} {\bibfnamefont {T.~L.}\ \bibnamefont
			{Curtright}}, \bibinfo {author} {\bibfnamefont {D.~B.}\ \bibnamefont
			{Fairlie}}, \ and\ \bibinfo {author} {\bibfnamefont {C.~K.}\ \bibnamefont
			{Zachos}},\ }\href@noop {} {\emph {\bibinfo {title} {{a Concise Treatise On
					Quantum Mechanics In Phase}}}}\ (\bibinfo  {publisher} {World Scientific},\
	\bibinfo {year} {2014})\BibitemShut {NoStop}%
	\bibitem [{\citenamefont {Groenewold}(1946)}]{Groenewold1946}%
	\BibitemOpen
	\bibfield  {author} {\bibinfo {author} {\bibfnamefont {H.~J.}\ \bibnamefont
			{Groenewold}},\ }\href {\doibase 10.1016/S0031-8914(46)80059-4} {\bibfield
		{journal} {\bibinfo  {journal} {Physica}\ }\textbf {\bibinfo {volume} {12}},\
		\bibinfo {pages} {405} (\bibinfo {year} {1946})}\BibitemShut {NoStop}%
	\bibitem [{\citenamefont {Akhiezer}\ and\ \citenamefont
		{Glazman}(1993)}]{LinearOperatorsHilbertSpace}%
	\BibitemOpen
	\bibfield  {author} {\bibinfo {author} {\bibfnamefont {N.~I.}\ \bibnamefont
			{Akhiezer}}\ and\ \bibinfo {author} {\bibfnamefont {I.~M.}\ \bibnamefont
			{Glazman}},\ }\href@noop {} {\emph {\bibinfo {title} {Theory of Linear
				Operators in Hilbert Space}}},\ \bibinfo {edition} {2nd}\ ed.\ (\bibinfo
	{publisher} {Dover Publications},\ \bibinfo {year} {1993})\BibitemShut
	{NoStop}%
	\bibitem [{\citenamefont {von Neumann}(1932{\natexlab{b}})}]{neumannbook}%
	\BibitemOpen
	\bibfield  {author} {\bibinfo {author} {\bibfnamefont {J.}~\bibnamefont {von
				Neumann}},\ }\href@noop {} {\emph {\bibinfo {title} {Mathematical Foundations
				of Quantum Mechanics}}}\ (\bibinfo  {publisher} {Princeton University
		press},\ \bibinfo {address} {Princeton},\ \bibinfo {year} {1932})\BibitemShut
	{NoStop}%
	\bibitem [{\citenamefont {Naimark}(2009)}]{naimark2009linear}%
	\BibitemOpen
	\bibfield  {author} {\bibinfo {author} {\bibfnamefont {M.~A.}\ \bibnamefont
			{Naimark}},\ }\href@noop {} {\emph {\bibinfo {title} {Linear differential
				operators}}}\ (\bibinfo  {publisher} {Dover Publications},\ \bibinfo
	{address} {Mineola, N.Y},\ \bibinfo {year} {2009})\BibitemShut {NoStop}%
	\bibitem [{Note1()}]{Note1}%
	\BibitemOpen
	\bibinfo {note} {In the quantum case, the trace of this operator is the
		von-Neumann entropy. This similarly obeys an $H$-theorem when $\protect
		\mathaccentV {hat}05E{\rho }$ is evolved by a Lindblad equation \protect
		\citep {Alicki}.}\BibitemShut {Stop}%
	\bibitem [{\citenamefont {Goldstein}(2014)}]{Goldstein}%
	\BibitemOpen
	\bibfield  {author} {\bibinfo {author} {\bibfnamefont {H.}~\bibnamefont
			{Goldstein}},\ }\href@noop {} {\emph {\bibinfo {title} {Classical
				Mechanics}}},\ \bibinfo {edition} {3rd}\ ed.\ (\bibinfo  {publisher}
	{Pearson},\ \bibinfo {address} {Essex},\ \bibinfo {year} {2014})\BibitemShut
	{NoStop}%
	\bibitem [{\citenamefont {Samuel}\ and\ \citenamefont
		{Bhandari}(1988)}]{Samuel1988}%
	\BibitemOpen
	\bibfield  {author} {\bibinfo {author} {\bibfnamefont {J.}~\bibnamefont
			{Samuel}}\ and\ \bibinfo {author} {\bibfnamefont {R.}~\bibnamefont
			{Bhandari}},\ }\href@noop {} {\bibfield  {journal} {\bibinfo  {journal}
			{Phys. Rev. Lett.}\ }\textbf {\bibinfo {volume} {60}},\ \bibinfo {pages}
		{2339} (\bibinfo {year} {1988})}\BibitemShut {NoStop}%
	\bibitem [{\citenamefont {Hannay}(1985)}]{HANNAY1985}%
	\BibitemOpen
	\bibfield  {author} {\bibinfo {author} {\bibfnamefont {J.~H.}\ \bibnamefont
			{Hannay}},\ }\href@noop {} {\bibfield  {journal} {\bibinfo  {journal} {J.
				Phys. A}\ }\textbf {\bibinfo {volume} {18}},\ \bibinfo {pages} {221}
		(\bibinfo {year} {1985})}\BibitemShut {NoStop}%
	\bibitem [{\citenamefont {Usatenko}, \citenamefont {Provost},\ and\
		\citenamefont {Vall{\'{e}}e}(1996)}]{0305-4470-29-10-035}%
	\BibitemOpen
	\bibfield  {author} {\bibinfo {author} {\bibfnamefont {O.~V.}\ \bibnamefont
			{Usatenko}}, \bibinfo {author} {\bibfnamefont {J.~P.}\ \bibnamefont
			{Provost}}, \ and\ \bibinfo {author} {\bibfnamefont {G.}~\bibnamefont
			{Vall{\'{e}}e}},\ }\href {http://stacks.iop.org/0305-4470/29/i=10/a=035}
	{\bibfield  {journal} {\bibinfo  {journal} {J. Phys. A}\ }\textbf {\bibinfo
			{volume} {29}},\ \bibinfo {pages} {2607} (\bibinfo {year}
		{1996})}\BibitemShut {NoStop}%
	\bibitem [{\citenamefont {Risken}(1989)}]{risken1989}%
	\BibitemOpen
	\bibfield  {author} {\bibinfo {author} {\bibfnamefont {H.}~\bibnamefont
			{Risken}},\ }\href@noop {} {\emph {\bibinfo {title} {The Fokker-Planck
				Equation : Methods of Solution and Applications}}}\ (\bibinfo  {publisher}
	{Springer Berlin Heidelberg},\ \bibinfo {address} {Berlin, Heidelberg},\
	\bibinfo {year} {1989})\ Chap.\ \bibinfo {chapter} {5.4}\BibitemShut
	{NoStop}%
	\bibitem [{\citenamefont {Case}(2008)}]{Case2008}%
	\BibitemOpen
	\bibfield  {author} {\bibinfo {author} {\bibfnamefont {W.~B.}\ \bibnamefont
			{Case}},\ }\href {\doibase 10.1119/1.2957889} {\bibfield  {journal} {\bibinfo
			{journal} {Am. J. Phys.}\ }\textbf {\bibinfo {volume} {76}},\ \bibinfo
		{pages} {937} (\bibinfo {year} {2008})},\ \Eprint
	{http://arxiv.org/abs/0512060} {arXiv:0512060} \BibitemShut {NoStop}%
	\bibitem [{\citenamefont {Bondar}\ \emph {et~al.}(2013)\citenamefont {Bondar},
		\citenamefont {Cabrera}, \citenamefont {Zhdanov},\ and\ \citenamefont
		{Rabitz}}]{Bondar2013a}%
	\BibitemOpen
	\bibfield  {author} {\bibinfo {author} {\bibfnamefont {D.~I.}\ \bibnamefont
			{Bondar}}, \bibinfo {author} {\bibfnamefont {R.}~\bibnamefont {Cabrera}},
		\bibinfo {author} {\bibfnamefont {D.~V.}\ \bibnamefont {Zhdanov}}, \ and\
		\bibinfo {author} {\bibfnamefont {H.~A.}\ \bibnamefont {Rabitz}},\ }\href
	{\doibase 10.1103/PhysRevA.88.052108} {\bibfield  {journal} {\bibinfo
			{journal} {Phys. Rev. A}\ }\textbf {\bibinfo {volume} {88}},\ \bibinfo
		{pages} {052108} (\bibinfo {year} {2013})},\ \Eprint
	{http://arxiv.org/abs/1202.3628} {arXiv:1202.3628} \BibitemShut {NoStop}%
	\bibitem [{\citenamefont {Griffiths}(2004)}]{Griffiths}%
	\BibitemOpen
	\bibfield  {author} {\bibinfo {author} {\bibfnamefont {D.~J.}\ \bibnamefont
			{Griffiths}},\ }\href
	{https://www.amazon.com/Introduction-Quantum-Mechanics-David-Griffiths/dp/0131118927?SubscriptionId=AKIAIOBINVZYXZQZ2U3A&tag=chimbori05-20&linkCode=xm2&camp=2025&creative=165953&creativeASIN=0131118927}
	{\emph {\bibinfo {title} {Introduction to Quantum Mechanics (2nd Edition)}}}\
	(\bibinfo  {publisher} {Pearson Prentice Hall},\ \bibinfo {year}
	{2004})\BibitemShut {NoStop}%
	\bibitem [{\citenamefont {Steuernagel}, \citenamefont {Kakofengitis},\ and\
		\citenamefont {Ritter}(2013)}]{WignerFlow}%
	\BibitemOpen
	\bibfield  {author} {\bibinfo {author} {\bibfnamefont {O.}~\bibnamefont
			{Steuernagel}}, \bibinfo {author} {\bibfnamefont {D.}~\bibnamefont
			{Kakofengitis}}, \ and\ \bibinfo {author} {\bibfnamefont {G.}~\bibnamefont
			{Ritter}},\ }\href {\doibase 10.1103/PhysRevLett.110.030401} {\bibfield
		{journal} {\bibinfo  {journal} {Phys. Rev. Lett.}\ }\textbf {\bibinfo
			{volume} {110}},\ \bibinfo {pages} {030401} (\bibinfo {year}
		{2013})}\BibitemShut {NoStop}%
	\bibitem [{\citenamefont {Oliva}, \citenamefont {Kakofengitis},\ and\
		\citenamefont {Steuernagel}(2018)}]{anharmonicwigner}%
	\BibitemOpen
	\bibfield  {author} {\bibinfo {author} {\bibfnamefont {M.}~\bibnamefont
			{Oliva}}, \bibinfo {author} {\bibfnamefont {D.}~\bibnamefont {Kakofengitis}},
		\ and\ \bibinfo {author} {\bibfnamefont {O.}~\bibnamefont {Steuernagel}},\
	}\href {\doibase https://doi.org/10.1016/j.physa.2017.10.047} {\bibfield
		{journal} {\bibinfo  {journal} {Physica A}\ }\textbf {\bibinfo {volume}
			{502}},\ \bibinfo {pages} {201 } (\bibinfo {year} {2018})}\BibitemShut
	{NoStop}%
	\bibitem [{\citenamefont {Jaffe}(2010)}]{Jaffe2010}%
	\BibitemOpen
	\bibfield  {author} {\bibinfo {author} {\bibfnamefont {R.~L.}\ \bibnamefont
			{Jaffe}},\ }\href {\doibase 10.1119/1.3298428} {\bibfield  {journal}
		{\bibinfo  {journal} {Am. J. Phys.}\ }\textbf {\bibinfo {volume} {78}},\
		\bibinfo {pages} {620} (\bibinfo {year} {2010})}\BibitemShut {NoStop}%
	\bibitem [{\citenamefont {Alicki}(1979)}]{Alicki}%
	\BibitemOpen
	\bibfield  {author} {\bibinfo {author} {\bibfnamefont {R.}~\bibnamefont
			{Alicki}},\ }\href {http://stacks.iop.org/0305-4470/12/i=5/a=007} {\bibfield
		{journal} {\bibinfo  {journal} {J. Phys. A}\ }\textbf {\bibinfo {volume}
			{12}},\ \bibinfo {pages} {L103} (\bibinfo {year} {1979})}\BibitemShut
	{NoStop}%
\end{thebibliography}
\addcontentsline{toc}{section}{\refname}

\end{document}